\documentclass[
  aip,
  pop,
  amsmath,amssymb,
  reprint,
  floatfix,
  superscriptaddress
]{revtex4-1}

\usepackage[utf8]{inputenc}
\usepackage[T1]{fontenc}
\usepackage{mathptmx}
\usepackage{graphicx}
\usepackage[usenames,dvipsnames]{xcolor}
\usepackage{booktabs}
\usepackage{hyperref}
\usepackage{url}
\usepackage{comment}
\usepackage{etoolbox}
\graphicspath{{figures/}}

\makeatletter
\def\@email#1#2{%
  \endgroup
  \patchcmd{\titleblock@produce}
    {\expandafter\produce@RRAP\expandafter{\@date}}
    {\produce@RRAP{*#1\href{mailto:#2}{#2}}\expandafter\produce@RRAP\expandafter{\@date}}
    {}{}%
}%
\makeatother

\newcommand{\pqls}{\textsc{PQLS}}
\newcommand{\tglf}{\textsc{TGLF}}
\newcommand{\cgyro}{\textsc{CGYRO}}
\newcommand{\gene}{\textsc{GENE}}
\newcommand{\qualikiz}{\textsc{QuaLiKiz}}
\newcommand{\numpyro}{\textsc{NumPyro}}
\newcommand{\jax}{\textsc{JAX}}

\newcommand{\sat}{\mathrm{SAT}}

\newcommand{\sattwo}{\sat 2}
\newcommand{\satthree}{\sat 3}
\newcommand{\qe}{Q_e}
\newcommand{\qi}{Q_i}
\newcommand{\gpart}{\Gamma_e}
\newcommand{\ky}{k_y}

\newcommand{\thetavec}{\boldsymbol{\theta}}
\newcommand{\nlpd}{\mathrm{NLPD}}
\newcommand{\qla}{\mathrm{QLA}}

\makeatletter
\newcommand{\dbloverline}[1]{\overline{\dbl@overline{#1}}}
\newcommand{\dbl@overline}[1]{\mathpalette\dbl@@overline{#1}}
\newcommand{\dbl@@overline}[2]{%
  \begingroup
  \sbox\z@{$\m@th#1\overline{#2}$}%
  \ht\z@=\dimexpr\ht\z@-2\dbl@adjust{#1}\relax
  \box\z@
  \ifx#1\scriptstyle\kern-\scriptspace\else
  \ifx#1\scriptscriptstyle\kern-\scriptspace\fi\fi
  \endgroup
}
\newcommand{\dblunderline}[1]{\@@underline{\dbl@underline{#1}}}
\newcommand{\dbl@underline}[1]{\mathpalette\dbl@@underline{#1}}
\newcommand{\dbl@@underline}[2]{%
  \begingroup
  \sbox\z@{$\m@th#1\@@underline{#2}$}%
  \dp\z@=\dimexpr\dp\z@-2\dbl@adjust{#1}\relax
  \box\z@
  \ifx#1\scriptstyle\kern-\scriptspace\else
  \ifx#1\scriptscriptstyle\kern-\scriptspace\fi\fi
  \endgroup
}
\newcommand{\dbl@adjust}[1]{%
  \fontdimen8
  \ifx#1\displaystyle\textfont\else
  \ifx#1\textstyle\textfont\else
  \ifx#1\scriptstyle\scriptfont\else
  \scriptscriptfont\fi\fi\fi 3
}
\makeatother

\renewcommand{\vec}[1]{{%
  \mspace{0.5mu}%
  \underline{\mspace{-0.5mu}#1_{}\kern-\scriptspace\mspace{-0.5mu}}%
  \mspace{0.5mu}%
  \mathcorr{#1}%
}}
\newcommand{\mat}[1]{{%
  \mspace{0.5mu}%
  \dblunderline{\mspace{-0.5mu}#1_{}\kern-\scriptspace\mspace{-0.5mu}}%
  \mspace{0.5mu}%
  \mathcorr{#1}%
}}
\makeatletter
\newcommand{\mathcorr}[1]{\mathpalette\math@corr{#1}}
\newcommand{\math@corr}[2]{%
  \begingroup
  \sbox\z@{$\m@th#1#2$}\sbox2{$\m@th#1#2_{}\kern-\scriptspace$}%
  \kern\dimexpr\wd\z@-\wd\tw@\relax
  \endgroup
}
\makeatother
\hypersetup{
  pdftitle={PQLS: A Quasilinear Gyrokinetic Transport Solver with a Bayesian Saturation-Rule Closure},
  pdfauthor={F. Wilms, A. Agrawal, J. J. Freigang, T. F. Neiser, B. J. Frei, and O. Meneghini},
  pdfkeywords={quasilinear transport, gyrokinetic turbulence, stellarator, saturation rule, Bayesian calibration, uncertainty quantification}
}

\begin{document}

\title{\pqls{}: A Quasilinear Gyrokinetic Transport Solver with a Bayesian Saturation-Rule Closure}

\author{F. Wilms}
\email{fwilms@proximafusion.com}
\thanks{These authors contributed equally to this work.}
\affiliation{Proxima Fusion, Flößergasse 2, Munich, 81369, Germany}

\author{A. Agrawal}
\thanks{These authors contributed equally to this work.}
\affiliation{Proxima Fusion, Flößergasse 2, Munich, 81369, Germany}

\author{J.~J. Freigang}
\affiliation{Proxima Fusion, Flößergasse 2, Munich, 81369, Germany}

\author{T.~F. Neiser}
\affiliation{General Atomics, P.O. Box 85608, San Diego, CA, 92186, United States of America}

\author{B.~J. Frei}
\affiliation{Max Planck Institute for Plasma Physics, Boltzmannstr. 2, 85748, Garching, Germany}

\author{O. Meneghini}
\affiliation{Proxima Fusion, Flößergasse 2, Munich, 81369, Germany}

\date{\today}

\begin{abstract}
Quasilinear models make gyrokinetic turbulent-transport predictions
sufficiently fast for integrated modelling, but their predictive capability
is limited by two factors: the physical and geometrical applicability of the linear
solver, and the validity of the saturation rule used to close the model. We present
the \emph{Predictive Quasilinear Solver} (\pqls{}), a quasilinear gyrokinetic
transport solver formulated in general magnetic geometry. Its implementation as
an eigenvalue solver retains electromagnetic and collisional effects, provides access to
dominant and subdominant modes and is differentiable with respect to all plasma
parameters. Linear benchmarks against \gene{} reproduce the growth rates,
frequencies, and eigenfunctions. We additionally formulate the saturation-rule closure as a Bayesian
inference problem that distinguishes uncertainty in its fitted coefficients
from the residual model-form uncertainty. The approach is demonstrated by calibrating
the $\satthree{}$ rule on \pqls{} quasilinear weights against published nonlinear
\cgyro{} cases. In addition to improving the robustness of the calibration,
the new method also quantifies the uncertainty in each of the fit coefficients.
Such uncertainty is propagated through transport calculations to produce error-aware profiles
that are compared to the ones obtained from the full gyrokinetic simulation, showing excellent agreement.
\end{abstract}


\maketitle

\section{Introduction}
\label{sec:introduction}
Reliable prediction of density and temperature profiles is essential for the
design, optimisation, and operation of magnetically confined fusion devices.
These profiles are strongly influenced by turbulent cross-field transport,
for which nonlinear gyrokinetic simulations provide the highest-fidelity
first-principles description currently available
\cite{dimits2000,jenko2000,candy2016}. Its computational cost, however, makes
it impractical to perform a nonlinear simulation at every radius and at every
iteration of integrated transport calculations used for scenario optimisation or
reactor-design workflow. Active-learning strategies reduce the computational cost by reducing
the number of gyrokinetic evaluations needed to solve the transport problem \cite{portals},
but the maximum speedup possible will still be limited by the time it takes to run a single gyrokinetic simulation.
On the other hand, machine-learning based approaches \cite{meneghini2017,vandeplassche2020}
have demonstrated the ability to speedup simulations by several orders of magnitude, but these
require so much data that is prohibitive to generate with direct gyrokinetic solvers.
Both approaches, would greatly benefit of physics-based reduced transport models to combine broad physical
applicability with sufficiently fast evaluation.

Quasilinear transport models occupy this intermediate level of fidelity. They
first solve the linear gyrokinetic problem over a prescribed wavenumber
spectrum and use the resulting eigenvalues and eigenfunctions to construct
quasilinear transport weights. A so-called saturation rule then estimates the
fluctuation amplitudes that would otherwise be determined by the nonlinear
dynamics and converts the linear response into turbulent particle and heat
fluxes.

The linear problem is generally solved in two ways so far, each with its own limitation.
On the one hand, high-fidelity gyrokinetic codes such as \cgyro \cite{candy2016} or \gene{} \cite{jenko2000,gene_code} retain the complete kinetic physics
and support nonaxisymmetric configurations, but their linear calculations
are commonly performed as initial-value simulations. The long-time evolution of such a calculation is
dominated by the fastest-growing eigenmode, making subdominant branches
difficult to access without specialised procedures. These branches can
nevertheless become dominant as the wavenumber or plasma parameters are
varied and are therefore important for robust transport calculations \cite{li2022assessing}.
On the other hand, models such as \tglf{} \cite{staebler2007,staebler2020} and
\qualikiz{} \cite{bourdelle2016} implement linear eigenvalue solvers that are much faster while retaining subdominant modes in their calculations
but make some important approximations (eg. gyrofluid and axisymmetric tokamak geometry).

Independently of how the linear gyrokinetic problem is solved, one needs to introduce an
empirical saturation rule to translate the linear growth rate into a saturated fluctuation intensity.
Such rules are constructed by choosing a
functional form and fitting its coefficients to a finite database of
nonlinear simulations \cite{staebler2021,dudding2022}.
This approach leads to two sources of uncertainties. First, the limited calibration database
does not determine the coefficients exactly, giving rise to parametric
uncertainty. Second, the assumed saturation-rule ansatz is only an
approximation to the nonlinear dynamics, giving rise to model-form
uncertainty that cannot be eliminated by retuning the same coefficients.
Conventional calibration produces a single set of coefficients and therefore leaves the uncertainties unreported.
When only a point estimate is passed to an integrated transport solver, both
uncertainties are not represented in the reported profile even though they remain
present in the underlying model.

In this work, we address these limitations through the development of
\emph{Predictive Quasilinear Solver} (\pqls{}). \pqls{} solves the linear gyrokinetic system
using a Hermite-Laguerre representation of velocity space based on the
flexible gyro-fluid formulation of \cite{staebler2023flexible}. It is posed
in general magnetic geometry, retains electrostatic and electromagnetic as well as collisional effects, and allows the velocity-space moment resolution to be selected
according to the physics of the instability. Since it is run as an eigensolver it can recover
both dominant and physically relevant subdominant eigenvalues and eigenvectors. We verify
the linear implementation against \gene{} for the Cyclone Base Case with
adiabatic and kinetic electrons and for a collisionless microtearing mode
benchmark.

We then formulate the saturation-rule closure as a Bayesian inference
problem, which provides a systematic
way to infer the coefficient distribution while representing the unresolved
model discrepancy explicitly
\cite{kennedy2001,brynjarsdottir2014,duraisamy2019}.
The methodology is demonstrated by calibrating the $\satthree{}$ saturation rule \cite{dudding2022} using \pqls{} linear simulations as well as published nonlinear \cgyro{} simulation counterparts. The inference determines a joint posterior over the
saturation-rule coefficients and discrepancy term parameters. The calibrated rule improves the
prediction of ion heat, electron heat, and electron particle fluxes relative
to the published coefficients.
Finally, we propagate the posterior through a steady-state transport
calculation. The resulting uncertainty band in the ion temperature profile envelops a high-fidelity reference profile computed using \gene{}, demonstrating how uncertainty in the
quasilinear closure can be carried into a downstream plasma prediction.

Such probabilistic calibration is not specific to \pqls{} or to
$\satthree{}$. Any quasilinear model in which linear transport weights are
combined with a parametrised saturation rule admits the same separation
between parametric and model-form uncertainty.

The remainder of the paper is organised as follows.
Sec.~\ref{sec:solver} introduces the \pqls{} model, eigensolver, linear
benchmarks, and quasilinear flux interface.
Sec.~\ref{sec:calibration} formulates the Bayesian calibration and
predictive uncertainty model.
Sec.~\ref{sec:results} applies the methodology to the $\satthree{}$
saturation rule using nonlinear \cgyro{} data.
Sec.~\ref{sec:profiles} propagates the calibrated closure through a
steady-state profile calculation, and Sec.~\ref{sec:conclusions}
summarises the results and discusses future extensions.

\section{The \pqls{} solver}
\label{sec:solver}
PQLS is a linear eigenvalue solver, designed to calculate the
linear modes characteristic with the same model fidelity of gyrokinetic flux-tube codes such as CGYRO and GENE, but at the fraction of the computational cost, and with the ability to easily calculate the subdominant modes in the system as well.

\subsection{Model description}
PQLS solves the linear gyrokinetic system \cite{hazeltine2003plasma} in
field-aligned Clebsch coordinates $(\rho_{\rm tor},y,z)$. The equilibrium
magnetic field is written as
\begin{equation}
  \vec{B}_0
  =
  \mathcal{C}(\rho_{\rm tor})
  \nabla\rho_{\rm tor}\times\nabla y ,
\end{equation}
where
$\rho_{\rm tor}=\sqrt{\Phi_{\rm tor}/\Phi_{\rm edge}}$ is the normalized
toroidal-flux radial coordinate and $\mathcal{C}(\rho_{\rm tor})$ is a geometry-dependent normalisation factor. The binormal coordinate $y$ is proportional to the
field-line label $\alpha$,
\begin{equation}
  y=\sigma_{B_{\rm p}}C_y\alpha,
  \qquad
  \alpha=q(\rho_{\rm tor})\theta^*-\phi ,
\end{equation}
where $\theta^*$ is a straight-field-line poloidal PEST angle \cite{PEST}, $\phi$ is the
geometrical toroidal angle, and $q$ is the safety factor. Equivalently,
$q=1/\iota$, where $\iota$ denotes the rotational transform. The factor
$C_y$ determines the normalisation of the binormal coordinate, while
$\sigma_{B_{\rm p}}$ denotes the sign of the poloidal magnetic field.
The field-line-following coordinate $z$ is taken to be a monotonic parameter along the magnetic field, oriented such that $\vec{b}\cdot\nabla z>0$ for $\vec{b}=\vec{B}_0/B_0$. Its precise parameterization depends on the numerical representation of ballooning space chosen within a simulation.

The local equilibrium geometry entering the gyrokinetic system is
specified by the magnetic-field strength $B_0(z)$, the Clebsch factor
$\mathcal{C}$, the spatial Jacobian $\sqrt{g}$, and the contravariant
metric coefficients
\begin{equation}
  g^{ij}=\nabla u^i\cdot\nabla u^j,
  \qquad
  u^i\in\{\rho_{\rm tor},y,z\}.
\end{equation}
This coordinate formulation applies to equilibria with nested flux
surfaces in both tokamaks and stellarators. Stellarator applications of
PQLS will be considered in a future publication.

Going to the flux-tube limit, PQLS employs a Fourier representation in the
two directions perpendicular to the magnetic field lines, characterised by
the radial and binormal wavenumbers $k_{\rm x}$ and $k_{\rm y}$.
Adopting the Fourier convention
$\exp[-i(k_{\rm x}x+k_{\rm y}y)]$, perpendicular derivatives become
\begin{equation}
  \frac{\partial}{\partial x}\rightarrow -ik_{\rm x},
  \qquad
  \frac{\partial}{\partial y}\rightarrow -ik_{\rm y}.
\end{equation}
Consequently, the perpendicular gradient transforms according to
\begin{equation}
  \nabla_{\perp}
  =
  \nabla x\,\frac{\partial}{\partial x}
  +\nabla y\,\frac{\partial}{\partial y}
  \rightarrow
  -i\left(k_{\rm x}\nabla x+k_{\rm y}\nabla y\right)
  \equiv -i\vec{k}_{\perp}.
\end{equation}
Together with the
field-line-following coordinate $z$, the parallel velocity $v_{||}=\vec{v}\cdot \vec{b}$,
and the perpendicular kinetic energy $e_{\perp}=m_{\sigma}v_{\perp}^2/2$,
these variables provide a five-dimensional representation of the
nonadiabatic part of the perturbed distribution function
$h_{\sigma}(k_{\rm x},k_{\rm y},z,v_{||},e_{\perp})$. Here,
$\sigma$ labels a species with charge $q_{\sigma}$, mass $m_{\sigma}$,
temperature $T_{\sigma}$, and density $n_{\sigma}$.

Assuming a time dependence proportional to $\exp(-i\omega t)$, the
linear gyrokinetic equation can be written as the generalised eigenvalue
problem
\begin{equation}
\begin{aligned}
  &-i v_{||}\nabla_{||}h_{\sigma}
  +\omega_{\rm d,\sigma}h_{\sigma}
  \\
  &\quad
  +i e_{\perp}\nabla_{||}\ln B_0
  \left(
      \frac{1}{m_{\sigma}}
      \frac{\partial h_{\sigma}}{\partial v_{||}}
      -v_{||}
      \frac{\partial h_{\sigma}}{\partial e_{\perp}}
  \right)
  \\
  &\quad
  -iC[h_{\sigma}]
  -\frac{q_{\sigma}\widetilde{\chi}_{\sigma}}{T_{\sigma}}
   F_{\rm M,\sigma}\omega_{*,\sigma}^{\rm T}
  =
  \omega
  \left(
      h_{\sigma}
      -\frac{q_{\sigma}\widetilde{\chi}_{\sigma}}{T_{\sigma}}
       F_{\rm M,\sigma}
  \right),
\end{aligned}
\label{eq:GK_equation}
\end{equation}
where $\nabla_{||}=\vec{b}\cdot\nabla$, $F_{\rm M,\sigma}$ is the equilibrium
Maxwellian, and $C$ denotes the linearized collision operator. The
gyroaveraged generalised potential is
\begin{equation}
\begin{aligned}
  \widetilde{\chi}_{\sigma}
  ={}&
  J_0\left(b_{\sigma}\sqrt{x_{\sigma}}\right)
  \left(
      \phi_1-\frac{v_{||}}{c}A_{1,||}
  \right)
  \\
  &+
  \sqrt{\frac{2e_{\perp}}{m_{\sigma}c}}\,
  \frac{J_1\left(b_{\sigma}\sqrt{x_{\sigma}}\right)}
       {k_{\perp}}
  B_{1,||},
\end{aligned}
\end{equation}
where $\phi_1$, $A_{1,||}$, and $B_{1,||}$ are, respectively, the
electrostatic-potential perturbation, the parallel vector-potential
perturbation, and the parallel magnetic-field perturbation. The argument of the Bessel functions are expressed in terms of the normalised perpendicular-energy coordinate $\hat{e}_{\perp}=e_{\perp}/T_{\sigma}$ and the dimensionless perpendicular wavevector $b_{\sigma}=k_{\perp}v_{{\rm th},\sigma}/\Omega_{\sigma}$, where $v_{{\rm th},\sigma}=\sqrt{2T_{\sigma}/m_{\sigma}}$ and $\Omega_{\sigma}=q_{\sigma}B_0/(m_{\sigma}c)$. The magnetic
drift frequency is the sum of the grad-$B$ and curvature-drift
contributions,
\begin{equation}
  \omega_{\rm d,\sigma}
  =
  \omega_{\nabla B,\sigma}
  +
  \omega_{\kappa,\sigma},
\end{equation}
with
\begin{equation}
  \begin{aligned}
      \omega_{\nabla B,\sigma}
      &=
      -\vec{k}_{\perp}\cdot\vec{v}_{\nabla B,\sigma}
      =
      -\frac{e_{\perp}}{m_{\sigma}\Omega_{\sigma}}\,
      \vec{k}_{\perp}\cdot
      \left(\vec{b}\times\nabla\ln B_0\right),\\
      \omega_{\kappa,\sigma}
      &=
      -\vec{k}_{\perp}\cdot\vec{v}_{\kappa,\sigma}
      =
      -\frac{v_{||}^2}{\Omega_{\sigma}}\,
      \vec{k}_{\perp}\cdot
      \left(\vec{b}\times\left[\vec{b}\cdot \nabla \vec{b}\right]\right).
  \end{aligned}
  \end{equation}
  The diamagnetic frequency appearing in Eq.~(Eq.~\ref{eq:GK_equation}) is
  \begin{equation}
      \omega_{*,\sigma}^{\rm T}
      =
      \omega_{n,\sigma}
      +
      \omega_{T,\sigma}
      \left(
          \hat{e}_{\perp}
          +\hat{v}_{||}^2
          -\frac{3}{2}
      \right),
  \end{equation}
  where we introduced the normalised parallel velocity $\hat{v}_{||}=v_{||}/v_{\rm th,\sigma}$ and
  \begin{equation}
  \begin{aligned}
      \omega_{n,\sigma}
      &=
      -\frac{cT_{\sigma}}{q_{\sigma}B_0}\,
      \vec{k}_{\perp}\cdot
      \left(\vec{b}\times\nabla\ln n_{\sigma}\right),\\
      \omega_{T,\sigma}
      &=
      -\frac{cT_{\sigma}}{q_{\sigma}B_0}\,
      \vec{k}_{\perp}\cdot
      \left(\vec{b}\times\nabla\ln T_{\sigma}\right).
  \end{aligned}
  \end{equation}

The gyrokinetic equations are closed self-consistently by quasineutrality
and the parallel and perpendicular components of Ampère's law:
\begin{equation}
\fontsize{8.3}{10}\selectfont
\begin{aligned}
  &\!\!\left(
      \frac{k_{\perp}^2}{4\pi}
      +\sum_{\sigma}\frac{q_{\sigma}^2n_{\sigma}}{T_{\sigma}}
  \right)\phi_1
  =
  2\pi\sum_{\sigma}\frac{q_{\sigma}}{m_{\sigma}}
  \int_{-\infty}^{\infty}\int_0^{\infty}
  J_0\left(b_{\sigma}\sqrt{x_{\sigma}}\right)
  h_{\sigma}\,dv_{||}\,de_{\perp},
  \\
  &k_{\perp}^2A_{1,||}
  =
  \frac{8\pi^2}{c}
  \sum_{\sigma}\frac{q_{\sigma}}{m_{\sigma}}
  \int_{-\infty}^{\infty}\int_0^{\infty}
  J_0\left(b_{\sigma}\sqrt{x_{\sigma}}\right)
  v_{||}h_{\sigma}\,dv_{||}\,de_{\perp},
  \\
  &B_{1,||}
  =
  -\frac{8\pi^2}{c}
  \sum_{\sigma}\frac{q_{\sigma}}{m_{\sigma}}
  \sqrt{\frac{2}{m_{\sigma}}}
  \int_{-\infty}^{\infty}\int_0^{\infty}
  \frac{J_1\left(b_{\sigma}\sqrt{x_{\sigma}}\right)}
       {k_{\perp}}
  h_{\sigma}\,dv_{||}\,de_{\perp}.
\end{aligned}
\label{eq:GK_field_equations}
\end{equation}

Following the successful demonstration of computational efficiency of codes like GX \cite{mandell2024gx} and \cite{staebler2023flexible} using a spectral representation in velocity space, we follow the approach of the latter and normalise the distribution function, as well as the gyrokinetic system of equations according to the conventions of \cite{gorler2009multiscale} and discretise the modified normalised distribution function $\hat{H}_{\sigma}=\pi \hat{h}_{\sigma}/\sqrt{\hat{f}_{0,\sigma}}$ using a spectral Hermite-Laguerre representaiton of the normalised velocity coordinates $\hat{v}_{||}$ and $\hat{e}_{\perp}$, where $\hat{f}_{0,\sigma}=\pi \hat{F}_{\rm M,\sigma}$:
\begin{equation}
\begin{aligned}
&\hat{H}_{\sigma}
=\sum\limits_{i_u=1}^{n_u}\sum\limits_{i_e=1}^{n_e}\,
  \hat{\mathcal{H}}_{\sigma,i_u,i_e}(\hat{z})
  P_{i_u}(\hat{v}_{||})P_{i_e}(\hat{e}_{\perp})\\
&P_{i_u}(\hat{v}_{||})
=\frac{p_{i_u}(\hat{v}_{||})}{\sqrt{\pi}}
  \exp\left(-\frac{\hat{v}_{||}^2}{2}\right)\\
&P_{i_e}(\hat{e}_{\perp})
=p_{i_e}(\hat{e}_{\perp})
  \exp\left(-\frac{\hat{e}_{\perp}}{2}\right).
\end{aligned}
\label{eq:spectral_basis}
\end{equation}
Here, $p_{i_u}(\hat{v}_{||})$ are normalised versions of the physicist's Hermite polynomials and $p_{i_e}(\hat{e}_{\perp})$ are normalised Laguerre polynomials, such that
\begin{equation}
\displaystyle
    \begin{aligned}
        \int\limits_{-\infty}^{\infty}\, P_{i_u}(\hat{v}_{||}) P_{j_u}(\hat{v}_{||})\, d\hat{v}_{||}
        &=\int\limits_{-\infty}^{\infty}\,
          \frac{p_{i_u}(\hat{v}_{||}) p_{j_u}(\hat{v}_{||})}{\pi}
          \exp\left(-\hat{v}_{||}^2\right)\, d\hat{v}_{||}=\delta_{i_u,j_u}\\
        \int\limits_0^{\infty}\, P_{i_e}(\hat{e}_{\perp}) P_{j_e}(\hat{e}_{\perp}) \, d\hat{e}_{\perp}
        &=\int\limits_0^{\infty}\,
          p_{i_e}(\hat{e}_{\perp}) p_{j_e}(\hat{e}_{\perp})
          \exp\left(-\hat{e}_{\perp}\right)\, d\hat{e}_{\perp}=\delta_{i_e,j_e}.
    \end{aligned}
\end{equation}

PQLS internally combines the field-line-following coordinate $z$ with the radial wavenumber $k_{\rm x}$ to form a unified one-dimensional representation of ballooning space, allowing for easy interchange of various types of boundary conditions, such as fully periodic, twist-and-shift \cite{beer1995field}, as well as generalised twist-and-shift \cite{martin2018parallel} boundaries. The abstract formulation with respect to the parallel coordinate, as well as the choice of $\hat{e}_{\perp}$ instead of $\hat{\mu}\hat{B}(z)/\hat{T}_{\sigma}$ as perpendicular velocity coordinate in the moment expansion enables a flexible choice of discretisation schemes. At present, PQLS supports a representation using Hermite polynomials in the same way as presented in \cite{staebler2023flexible}, as well as a grid-based representation using finite difference methods employed by most gyrokinetic flux-tube codes to date, though other formulations like splines or Meixner–Pollaczek polynomials could be implemented easily in the future.

Collisional effects can be modeled either using a Dougherty operator \cite{francisquez2022improved}, or using Coulomb- or Sugama-type collision models without finite Larmor radius effects shown in \cite{frei2021development,frei2022numerical} in order to avoid catastrophic cancellation effects arising if a naive quadrature-based projection of the collision operators was used.

\subsection{Linear eigenvalue solve}
\label{sec:solver:linear}
By representing the individual coefficients of the eigenfunction representation (Eq.~\ref{eq:spectral_basis}) into a vector and representing the gyrokinetic potential $\chi$ through the distribution functions using the field equations (Eq.~\ref{eq:GK_field_equations}), the normalised spectral form of the gyrokinetic equation (Eq.~\ref{eq:GK_equation}) can be cast into a generalised eigenvalue problem
\begin{equation}
    \mat{\hat{A}}\, \vec{\hat{\mathcal{H}}}=\hat{\omega} \mat{\hat{B}}\,\vec{\hat{\mathcal{H}}},
\end{equation}
where the matrix $\mat{\hat{A}}$ represents the left-hand side and the matrix $\mat{\hat{B}}$ the right-hand side of (Eq.~\ref{eq:GK_equation}), respectively.
The complex eigenvalue $\hat{\omega}=\hat{\omega}_r+i\hat{\gamma}$ then gives the normalised real frequency and growth rate of the mode under consideration. While PQLS provides dense eigenvalue solvers in principle, production run cases at high resolutions will be run using a sparse Arnoldi solve to compute the most unstable eigenvalues. While plain Arnoldi iterations will converge to the eigenvalues of largest $|\hat{\omega}|$ - oftentimes strongly damped high-frequency modes - shift-invert methods combined with Arnoldi solves will converge to the eigenvalues nearest a target $\sigma$ that must be supplied in
advance. The latter is the more dangerous of the two in a parameter scan, since
a misplaced $\sigma$ does not fail but returns a well-converged
subdominant branch that is indistinguishable from the dominant one
without an independent check. Instead, we compute the eigenvalues of the generalised Cayley transform operator
 \begin{equation}
    \begin{aligned}
    \mat{\hat{C}}
    &=\underbrace{(\mat{\hat{A}}-\beta\mat{\hat{B}})^{-1}}_\text{$\equiv \mat{\hat{M}}^{-1}$}
      (\mat{\hat{A}}-\alpha \mat{\hat{B}})\\
    &=\mat{\hat{M}}^{-1} (\mat{I} + (\beta-\alpha)\mat{\hat{B}});\\
    \mu&=\frac{\hat{\omega}-\alpha}{\hat{\omega}-\beta},\\
    \hat{\omega}&=\frac{\mu\beta-\alpha}{\mu-1},
    \end{aligned}
    \label{eq:cayley}
  \end{equation}
  which shares every eigenvector of the pencil, remaps its eigenvalues invertibly,
  and requires no target $\sigma$. The shifts are constructed from a lower growth-rate threshold
  $\gamma_{\rm c}$ and a frequency window of centre $\omega_{\rm c}$ and
  half-width $\omega_{\rm w}$ as $\alpha=\omega_{\rm c}-\mathrm{i}\tau$,
  $\beta=\bar\alpha+2\mathrm{i}\gamma_{\rm c}$ with $\tau=\omega_{\rm w}/r$ - $r$ being a Bernstein-ellipse estimate of the
  range over which a Krylov iteration retains discrimination- , giving
  \begin{equation}
    |\mu|^{2}-1
    =\frac{4(\hat{\gamma}-\gamma_{\rm c})(\tau+\gamma_{\rm c})}
          {(\hat{\omega}_r-\omega_{\rm c})^{2}+(\hat{\gamma}-\tau-2\gamma_{\rm c})^{2}} .
    \label{eq:mu2}
  \end{equation}
  For $\tau+\gamma_{\rm c}>0$ - the single admissibility condition, equivalent
  to $\operatorname{Im}\beta>\operatorname{Im}\alpha$ - the sign of $|\mu|-1$ is
  that of $\hat{\gamma}-\gamma_{\rm c}$ at every frequency, so modes above the threshold
  are mapped outside the unit disc and are amplified by a largest-magnitude
  iteration while those below are suppressed; the amplification itself decays
  quadratically in $\hat{\omega}_r-\omega_{\rm c}$, restricting the search to a window of
  half-width $\omega_{\rm w}$. Since $|\mu|$-ordering under the Cayley transform demotes strongly growing modes far from the frequency window, we rank the leading $N$ Ritz pairs of the Schur form by the back-transformed growth rate and leave the thick-restart tail in $|\mu|$-ordering inside of the Arnoldi solver, ensuring that the modes we are interested in indeed are the most unstable ones.

\subsection{Linear benchmarks}
Before constructing quasilinear fluxes, we validate the linear eigenvalue
solver against corresponding \gene{} \cite{jenko2000} calculations. The benchmark suite
progressively tests ion temperature gradient modes (ITG) with adiabatic electrons,
coupled ion temperature gradient and trapped electron mode (TEM) branches when treating electrons kinetically, and collisionless electromagnetic microtearing modes (MTM). For
each case, we scan the binormal wavenumber $k_y$ and compare the growth rate
$\gamma$, the real mode frequency $\omega_r$, and selected field
eigenfunctions.

The spatial discretisation is characterised by $n_{k_x}$, the number of radial
Fourier harmonics in the ballooning representation, and by $n_z$, the number
of grid points along the field-aligned coordinate $z$. In the \gene{}
calculations, velocity space is discretised using $n_{v_\parallel}$ grid
points in the parallel velocity $v_\parallel$ and $n_\mu$ grid points in the
magnetic moment $\mu$. Throughout this paper, the corresponding
velocity-space domain used in \gene{} is
$(L_{v_\parallel},L_\mu)
=
\left(3v_{{\rm th},\sigma},9T_\sigma/B\right).
$
We therefore denote the complete \gene{} resolution by
$(n_{k_x},n_z,n_{v_\parallel},n_\mu)$.

In \pqls{}, velocity space is instead represented spectrally. We retain
$n_u$ Hermite moments in the normalised parallel velocity
$\hat{v}_\parallel$ and $n_e$ Laguerre moments in the normalised perpendicular
energy $\hat{e}_\perp$. The \pqls{} velocity-space resolution is consequently
specified by the moment pair $(n_u,n_e)$. In each benchmark, the
two codes use the same values of $n_{k_x}$ and $n_z$, while
$\left(n_u,n_e\right)$ is varied to assess convergence of the \pqls{}
velocity-space representation.

\subsubsection{Cyclone Base Case with adiabatic electrons}
\label{subsubsec:CBC_AE}
We first consider the well-established Cyclone Base Case (CBC), following the
benchmark specification described, for example, in
\cite{gorler2016intercode}. The calculation is performed at
$\rho_{\rm tor}=0.5$ in circular tokamak geometry, with inverse aspect ratio
$a/R=0.36$, safety factor $q_0=1.41$, and magnetic shear
$\hat{s}=0.8496$. Electrons are treated adiabatically: their perturbed density
is prescribed by a Boltzmann response, so that only the corresponding
adiabatic contribution is retained on the left-hand side of the
quasineutrality equation in (Eq.~\ref{eq:GK_field_equations}). We assume equal ion and electron temperatures $T_{\rm e}=T_{\rm i}$, as well as an ion temperature gradient of $R/L_{\rm T_{\rm i}}=6.96$ and a density gradient of $R/L_{\rm n}=2.23$.

\begin{figure}[htbp!]
  \centering
  \includegraphics[width=0.8\linewidth]{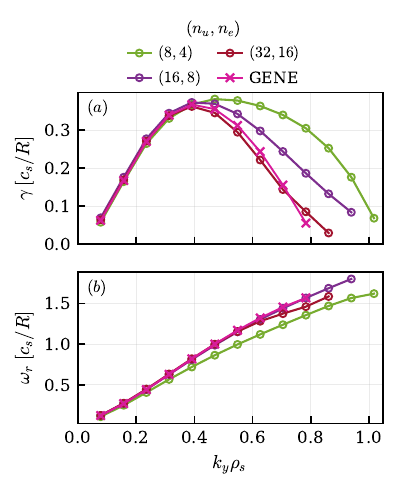}
  \caption{Linear (a) growth rate $\gamma$ and (b) real mode frequency
  $\omega_r$, in units of $c_s/R$, as functions of $k_y\rho_s$ for the
  Cyclone Base Case with adiabatic electrons. The \gene{} results (crosses) are
  compared with \pqls{} calculations (open circles) at moment resolutions
  $(n_u,n_e)=(8,4)$, $(16,8)$, and $(32,16)$.}
  \label{fig:CBC_eigenvalues_AE}
\end{figure}

Simulations done with \gene{} use the resolution$(n_{k_x},n_z,n_{v_\parallel},n_\mu)
=(32,32,32,9).$
The \pqls{} calculations use the same spatial resolution,
$\left(n_{k_x},n_z\right)=(32,32)$, while the moment resolution is increased
from $\left(n_u,n_e\right)=(8,4)$ to $(32,16)$. As shown in
Fig.~\ref{fig:CBC_eigenvalues_AE}, the \pqls{} growth rates and real
frequencies converge towards the \gene{} results as the number of retained
moments is increased. The converged calculation reproduces the location and magnitude of
the maximum growth rate particularly well, with the largest remaining
differences occurring at the highest wavenumbers.

\begin{figure}[htbp]
  \centering
  \includegraphics[width=0.75\linewidth]{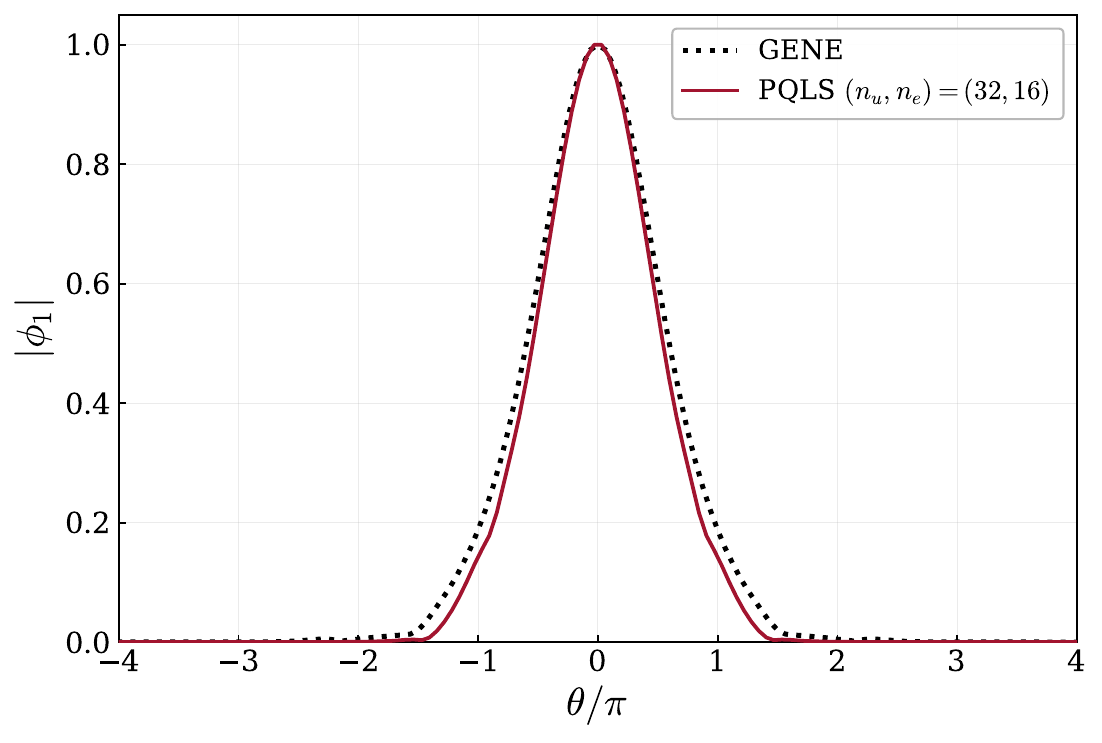}
   \caption{Normalised magnitude of the electrostatic-potential eigenfunction
  $|\phi_1(\theta)|$ for the most unstable mode at $k_y\rho_s=0.391$ in the
  Cyclone Base Case with adiabatic electrons. The \gene{} result (black dotted
  line) is compared with \pqls{} at
  $(n_u,n_e)=(32,16)$ (solid red line). Each eigenfunction is
  normalised to unit peak magnitude; $\theta=0$ denotes the outboard midplane.}

  \label{fig:AE_eigenfunction}
\end{figure}

Agreement in the eigenvalues is accompanied by agreement in the spatial mode
structure. Fig.~\ref{fig:AE_eigenfunction} compares the electrostatic
potential eigenfunction of the most unstable mode at
$k_y\rho_s=0.391$. Both calculations predict a mode localised around the
outboard midplane, $\theta=0$, with closely matching ballooning-space
structures.

\subsubsection{Electrostatic Cyclone Base Case with kinetic electrons}

We next retain the equilibrium and ion parameters of
Sec.~\ref{subsubsec:CBC_AE}, but treat both ions and electrons kinetically.
The calculation remains electrostatic. In addition to the gradients of the
adiabatic-electron case, we set the electron temperature gradient to
$R/L_{T_{\rm e}}=6.96$ and use the physical electron-to-ion mass ratio
$m_{\rm e}/m_{\rm i}=5.44\times10^{-4}$.

Resolving the electron dynamics requires a finer velocity-space grid in
\gene{}. We therefore use ($n_{k_x},n_z,n_{v_\parallel},n_\mu)
=
(32,32,64,32).
$
The \pqls{} calculations employ the same spatial resolution,
$(n_{k_x},n_z)=(32,32)$, and moment resolutions ranging from
$(n_u,n_e)=(16,8)$ to $(64,16)$.

\begin{figure}[htbp!]
  \centering
  \includegraphics[width=0.8\linewidth]{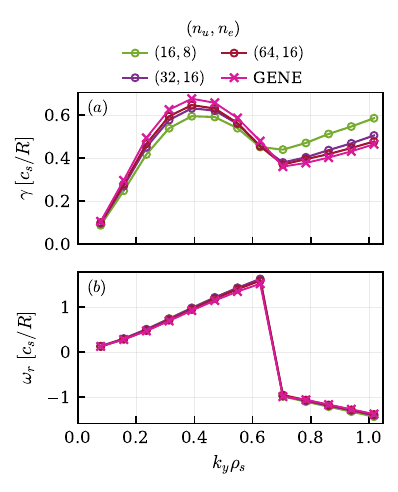}
  \caption{Dominant linear (a) growth rate $\gamma$ and (b) real mode frequency
  $\omega_r$, in units of $c_s/R$, as functions of $k_y\rho_s$ for the
  electrostatic Cyclone Base Case with kinetic electrons. The \gene{} results
  (crosses) are compared with \pqls{} calculations (open circles) at moment
  resolutions $(n_u,n_e)=(16,8)$, $(32,16)$, and $(64,16)$. The
  change in the sign of $\omega_r$ near $k_y\rho_s\simeq0.65$ corresponds to
  the transition from the ITG branch to the TEM branch.}
\label{fig:CBC_eigenvalues_KE}
\end{figure}

Fig.~\ref{fig:CBC_eigenvalues_KE} compares the dominant growth rate and real
frequency predicted by the two codes. A resolution of
$(n_u,n_e)=(32,16)$ is sufficient to describe the ITG branch at lower $k_y\rho_s$. At higher
wavenumbers, however, convergence of the TEM branch
requires up to $(n_u,n_e)=(64,16)$. At approximately
$k_y\rho_s\simeq0.65$, the change in the sign of $\omega_r$ marks the
transition from the ITG branch to the TEM branch as the dominant instability.
At the highest moment resolution, \pqls{} closely reproduces both branches of
the dominant \gene{} spectrum.

\begin{figure*}[htbp!]
  \centering
  \includegraphics[width=0.8\linewidth]{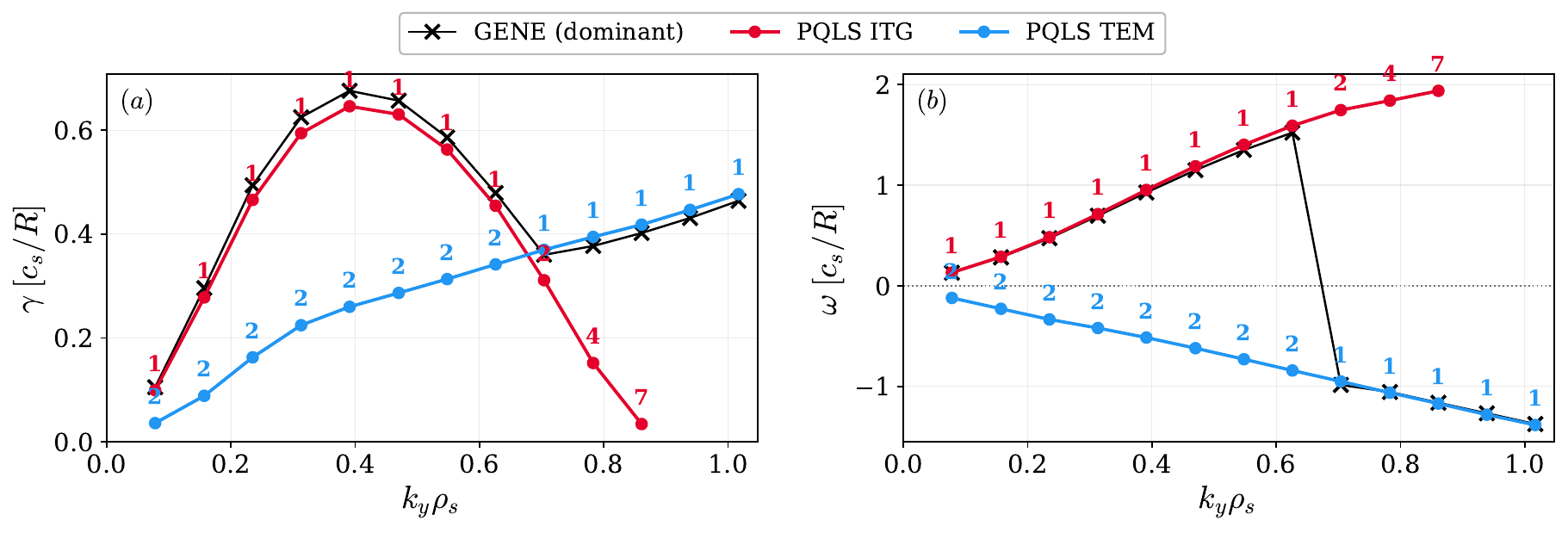}
  \caption{Tracked ITG (red) and TEM (blue) branches obtained with \pqls{},
  compared with the dominant \gene{} eigenvalue (black): (a) growth rate
  $\gamma$ and (b) real mode frequency $\omega_r$, both in units of $c_s/R$.
  The integer beside each \pqls{} point gives the rank of that eigenmode when
  the spectrum is ordered by decreasing growth rate; rank 1 denotes the
  dominant mode. The ITG branch is dominant at lower $k_y\rho_s$, whereas the
  TEM branch becomes dominant above $k_y\rho_s\simeq0.7$.}
  \label{fig:CBC_branches_KE}
\end{figure*}

The dominant-mode spectrum alone does not show how the two branches coexist
through this transition. Fig.~\ref{fig:CBC_branches_KE} therefore tracks the
most unstable ITG and TEM modes obtained with \pqls{} separately.

The ITG mode
is dominant at lower wavenumbers, while the TEM initially appears as the
second-most-unstable eigenmode. Their ordering reverses near $k_y\rho_s\simeq0.7$.
Beyond this point, the ITG branch remains identifiable as a
subdominant mode and approaches marginal stability at higher $k_y\rho_s$.
This branch tracking also illustrates the ability of the \pqls{} eigensolver
to recover physically relevant subdominant modes.

\begin{figure*}[htbp!]
  \centering
  \includegraphics[width=0.8\linewidth]{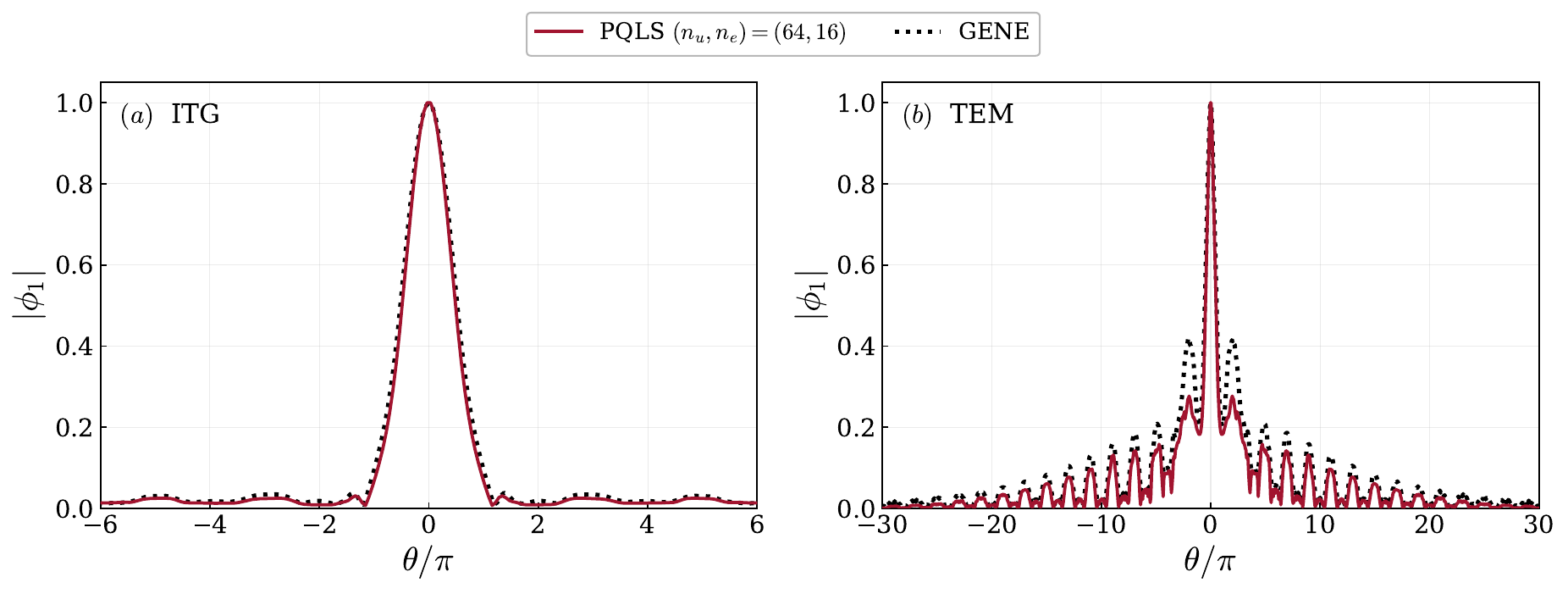}
    \caption{Normalised magnitude of the electrostatic-potential eigenfunction
  $|\phi_1(\theta)|$ for (a) the most unstable ITG mode at
  $k_y\rho_s=0.391$ and (b) the dominant TEM at $k_y\rho_s=1.017$ in the
  Cyclone Base Case with kinetic electrons. The \gene{} results (black dotted
  lines) are compared with the \pqls{} results at
  $(n_u,n_e)=(64,16)$ (solid red lines). Each eigenfunction is
  normalised to unit peak magnitude.}
  \label{fig:CBC_eigenfunctions_KE}
\end{figure*}

Finally, Fig.~\ref{fig:CBC_eigenfunctions_KE} compares the electrostatic
potential eigenfunctions for the most unstable ITG mode at
$k_y\rho_s=0.391$ and the dominant TEM at $k_y\rho_s=1.017$. The ITG
eigenfunctions obtained with \pqls{} and \gene{} are in close agreement. For
the more extended TEM eigenfunction, both codes predict strong localisation
around $\theta=0$ and the same oscillatory ballooning-space structure,
although some differences remain in the amplitudes of the subsidiary peaks.

\subsubsection{Collisionless microtearing modes in \texorpdfstring{$s-\alpha$}{s-alpha} geometry}
As a final linear benchmark, we consider the collisionless microtearing mode case introduced in \cite{frei2023moment}. In contrast to the
electrostatic Cyclone Base Case benchmarks, this case tests the electromagnetic
response of the solver, in particular the coupling to the parallel
vector-potential perturbation $A_{1,\parallel}$.

The benchmark employs an $s-$$\alpha$ equilibrium with magnetic shear
$\hat{s}=2.4$, $\alpha_{\rm MHD}=0$, safety factor $q=4$, and inverse aspect
ratio $a/R=0.18$. The calculation is performed at $\rho_{\rm tor}=0.5$ and
uses the reduced electron-to-ion mass ratio$ \sqrt{m_{\rm e}/m_{\rm i}}=1/19.24$.
The normalised equilibrium gradients are
$R/L_{T_{\rm i}}=0$, $R/L_{T_{\rm e}}=8$, and $R/L_n=3$, and the electron
plasma beta is $\beta_{\rm e}=0.02$. The strongly localised ballooning-space structure is resolved using $(n_{k_x},n_z)=(11,64)$ in both codes. Following the reference
benchmark \cite{frei2023moment}, the \gene{} calculation uses the complete
resolution $(n_{k_x},n_z,n_{v_\parallel},n_\mu)
=
(11,64,128,24)$.
The \pqls{} calculations use the same spatial resolution, while the
velocity-space moment resolution $(n_u,n_e)$ is varied to assess
convergence.

\begin{figure}[htbp!]
  \centering
  \includegraphics[width=0.8\linewidth]{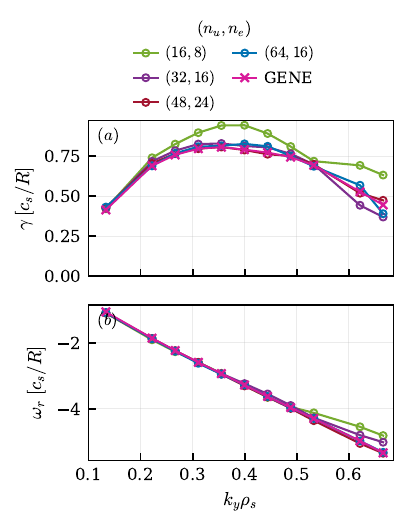}
  \caption{Dominant linear (a) growth rate $\gamma$ and (b) real mode frequency
  $\omega_r$, in units of $c_s/R$, as functions of $k_y\rho_s$ for the
  collisionless microtearing-mode benchmark. The \gene{} results (crosses) are
  compared with \pqls{} calculations (open circles) at moment resolutions
  $(n_u,n_e)=(16,8)$, $(32,16)$, $(48,24)$, and $(64,16)$. The
  comparison between $(48,24)$ and $(64,16)$ illustrates the importance of
  convergence in both the Hermite and Laguerre expansions.}
  \label{fig:MTM_eigenvalues}
\end{figure}

Fig.~\ref{fig:MTM_eigenvalues} compares the dominant growth rates and real
frequencies obtained with \pqls{} and \gene{}. At lower wavenumbers, a moment
resolution of $(n_u,n_e)=(32,16)$ already gives close agreement
with the \gene{} reference. Convergence over the full wavenumber range,
particularly at the largest values of $k_y\rho_s$, requires up to
$(n_u,n_e)=(48,24)$. The result at $(64,16)$ demonstrates that
increasing the number of parallel-velocity Hermite moments alone is
insufficient: adequate resolution in the perpendicular-energy Laguerre
expansion is also required.

In addition to the eigenvalues, we compare the electromagnetic mode structure
at $k_y\rho_s=0.3552$.

\begin{figure*}[htbp!]
  \centering
  \includegraphics[width=0.8\linewidth]{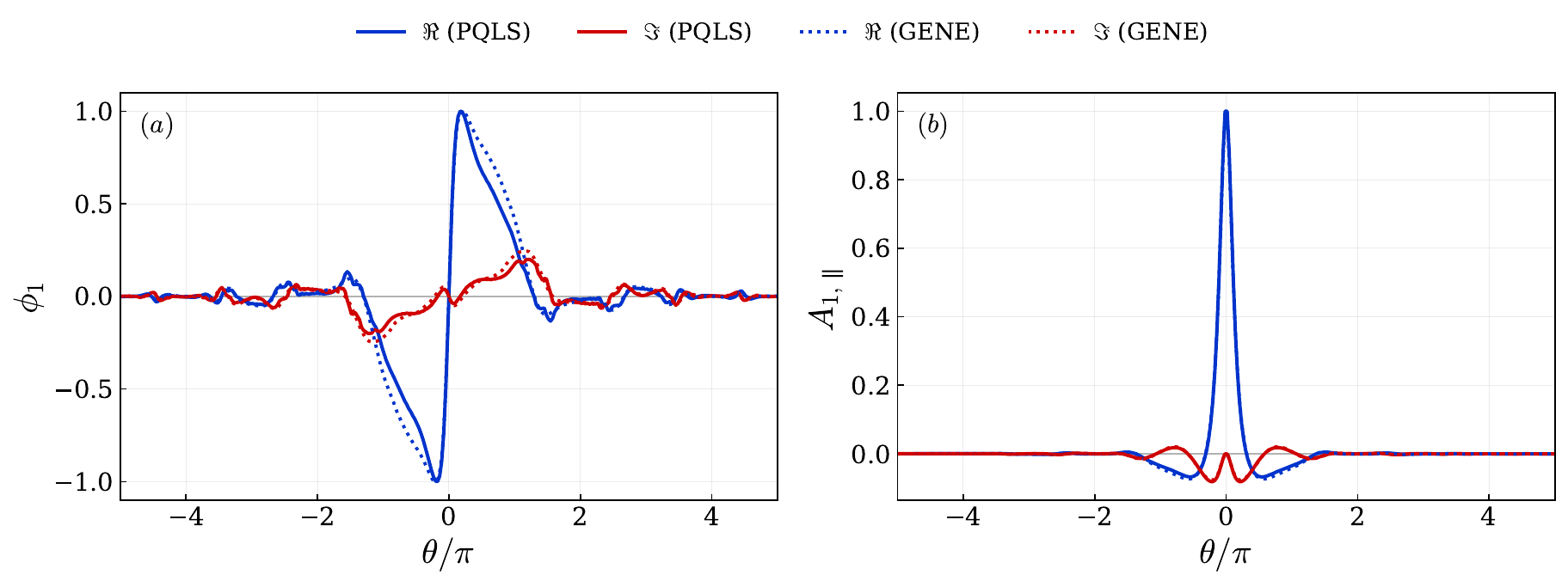}
 \caption{Real (blue) and imaginary (red) parts of the normalised
  (a) electrostatic potential $\phi_1$ and (b) parallel vector potential
  $A_{1,\parallel}$ at $k_y\rho_s=0.3552$ for the collisionless
  microtearing mode. The \pqls{} result at
  $(n_u,n_e)=(48,24)$ (solid lines) is compared with the \gene{}
  reference (dotted lines). The eigenfunctions exhibit the characteristic
  tearing parity of $\phi_1$ and ballooning parity of $A_{1,\parallel}$ about
  $\theta=0$.}
  \label{fig:MTM_eigenfunctions}
\end{figure*}

Fig.~\ref{fig:MTM_eigenfunctions} shows the real and
imaginary parts of the electrostatic potential $\phi_1$ and parallel vector
potential $A_{1,\parallel}$ in ballooning space. The \pqls{} calculation
reproduces both the localisation of the eigenfunctions and their characteristic
parities. In particular, the dominant component of $\phi_1$ has tearing
parity, whereas the dominant component of $A_{1,\parallel}$ has ballooning
parity about $\theta=0$. The agreement with \gene{} is especially close for
the parallel vector potential.

\subsection{Quasilinear flux and the saturation interface}
\label{sec:solver:ql}
Given an eigenvector $\vec{\hat{\mathcal{H}}}$ for a specific $\hat{k}_{\rm y}$, the gyrokinetic potential $\chi$ is calculated via the field equations (see Eq.~\ref{eq:GK_field_equations}), normalised to Gyrobohm units, of the form \cite{staebler2023flexible}
\begin{equation}
    \begin{aligned}
        &\mathrm{\textbf{Particle\, flux:}}\\
        &\hat{\Gamma}_{\sigma}^{\rm QL}(\hat{k}_{\rm y})={}\\[-0.6ex]
        &\qquad \hat{k}_{\rm y}\hat{n}_{0,\sigma}\Re\left\{i\left\langle  \int\limits_{-\infty}^{\infty} \int\limits_{0}^{\infty}\, \hat{H}_{\sigma}\hat{\tilde{\chi}}_{\sigma}\, d\hat{v}_{||} d\hat{e}_{\perp}\right\rangle_{\rm FS}\right\}\\
        &\mathrm{\textbf{Heat\, flux:}}\\
        &\hat{Q}_{\sigma}^{\rm QL}(\hat{k}_{\rm y})={}\\[-0.6ex]
        &\qquad \hat{k}_{\rm y}\hat{n}_{0,\sigma} \hat{T}_{0,\sigma}\Re\left\{i\left\langle  \int\limits_{-\infty}^{\infty} \int\limits_{0}^{\infty}\, (\hat{v}_{||}^2+\hat{e}_{\perp}) \hat{H}_{\sigma}\hat{\tilde{\chi}}_{\sigma}\, d\hat{v}_{||} d\hat{e}_{\perp}\right\rangle_{\rm FS}\right\},\\
    \end{aligned}
    \label{eq:QL_fluxes}
\end{equation}
where the angular brackets denote the flux-surface average and, compared with gyrokinetic literature, a minus sign is absent due to our ansatz of quantities of interest being proportional to $\exp(-i \textbf{k}_{\perp}\cdot \textbf{r})$. However, as all eigenvectors can only be determined up to a constant, only the ratios between different flux channels at a fixed $\hat{k}_{\rm y}$ is reliable right away, and we need a map to transform these results into an actionable flux spectrum that can be compared against first-principle turbulence simulations.

\pqls{} shares the standard quasilinear factorisation \cite{dudding2022}, which is what makes the
calibration of Sec.~\ref{sec:calibration} possible. The arbitrary eigenvector
amplitude noted above is first divided out: writing $I_{\phi}(\ky)$ for the
potential intensity $\hat{\phi_1}^{\dagger}\hat{\phi_1}$ of the same eigenmode, the
quasilinear weight of channel $m$ is
\begin{equation}
  \label{eq:ql_weight}
  w_m(\ky) \;=\; \frac{\hat{\mathcal{F}}^{\rm QL}_m(\ky)}{I_{\phi}(\ky)},
  \qquad
  m \in \bigl\{ (Q,i),\ (Q,e),\ (\Gamma,e) \bigr\},
\end{equation}
in which $\hat{\mathcal{F}}^{\rm QL}_m$ is the entry of (Eq.~\ref{eq:QL_fluxes})
belonging to channel $m$, that is $\hat{Q}^{\rm QL}_{\sigma}$ for an energy channel
and $\hat{\Gamma}^{\rm QL}_{\sigma}$ for a particle channel. Being a ratio, $w_m$
is independent of the eigenvector normalisation. The saturation rule then restores the physical fluctuation
level as the weighted sum over the
wavenumber grid,
\begin{equation}
  \label{eq:ql_sum}
  \mathcal{F}_m(\thetavec) \;=\; \sum_{\ky} w_m(\ky)\, I(\ky; \thetavec),
\end{equation}
with $I(\ky; \thetavec)$ being the saturated fluctuation intensity, and
empirical coefficient vector $\thetavec$, calibrated against the nonlinear reference data. The two
intensities are the same quantity taken at different stages: $I_{\phi}$ is the
linear eigenmode's own $|\phi_1|^{2}$, fixed only up to the arbitrary normalisation,
whereas $I(\ky;\thetavec)$ is the saturated $|\phi_1|^{2}$ the rule predicts.
Dividing by the first and multiplying by the second is what the quasilinear closure
amounts to.

\subsection{Software details}
\label{sec:solver:implementation}
PQLS is written in Julia, allowing for just in time compilation and efficient code performance. In addition, the implementation in Julia easily allows all outputs of the code - eigenvalues, eigenvectors and the quasilinear fluxes - to be differentiable with respect to all plasma parameters and wavenumbers and the calibration coefficients of the saturation rules. The code runs on CPU as well as on GPU, an thin interface to Python is provided in order to simplify incorporating the tool into other software frameworks. \pqls{} is available under an APACHE-2.0 license at
\url{https://github.com/proximafusion/PQLS.jl}.

\section{Bayesian calibration of the saturation rule}
\label{sec:calibration}
A quasilinear calculation does not determine the saturated fluctuation
intensity required to predict turbulent transport. An empirical saturation
rule supplies this closure through a prescribed functional form and
coefficients calibrated against nonlinear simulations
\cite{ahmed2021,sanderse2024,agrawal2024probabilistic}.
This construction introduces two sources of uncertainty: parametric
uncertainty arising from the limited information available to constrain
the coefficients, and model-form uncertainty associated with limitations
of the closure representation. Both are epistemic in origin
\cite{briggs2016,smith2013}, although reducing them may require different
approaches: additional informative reference data for the former, and
improvements to the closure representation for the latter.

The existing saturation rules report point estimates alone, for instance both $\sattwo$ \cite{staebler2021}
and $\satthree$ \cite{dudding2022} rule fits are determined via
a few tens of nonlinear cases. A point estimate
propagates into whatever downstream task consumes the rule, and the answer that task
returns can be confidently wrong. Both errors are therefore worth reporting
alongside the prediction, and Bayesian inference provides a framework for doing
so
\cite{jaynes2003,koutsourelakis2009,agrawal2025thesis}. The coefficients are estimated
through a posterior distribution rather than summarised only by a point estimate, which quantifies what a small calibration set
cannot fix, and the residual the ansatz cannot represent is carried explicitly as a
discrepancy term rather than absorbed into the coefficients
\cite{kennedy2001,agrawal2024probabilistic}. We infer both jointly and propagate
the resulting uncertainty through a transport calculation. It is carried out here for \pqls{} and
demonstrated on $\satthree$, which supplies the closure throughout, but the
proposed paradigm is not specific to either. \tglf{}
\cite{staebler2007,staebler2020}, \qualikiz{} \cite{bourdelle2016} and other
quasilinear transport models along with saturation rule can benefit from
the same treatment.

Three consequences of (Eq.~\ref{eq:ql_sum}) are relevant for the calibration. First, the linear solve giving $w_m$ is independent of $\thetavec$, so it is computed
once per case and cached, and only the algebraic saturation rule runs inside the
inference loop; the cost of calibration is therefore decoupled from the cost of
the solver. Second, the factorisation separates two error sources that are easily
conflated. An error in $w_m(\ky)$ belongs to the linear solver and cannot be
repaired by any choice of $\thetavec$, whereas an error in $I(\ky;\thetavec)$
belongs to the saturation rule and can be; Sec.~\ref{sec:results} separates the two
in practice. Third, any calibration procedure benefits from gradient of predicted fluxed w.r.t the coefficient vector i.e., 
$\nabla_{\thetavec}\mathcal{F}_m(\thetavec)$, and the present factorisation enables
it as $I(\ky;\thetavec)$ is algebraic and $w_m(k_y)$ are $\thetavec$ independent.

\subsection{Likelihood, priors and posterior} Indexing reference cases by $c$ and transport channels
by $m$, errors in a saturated flux are treated as multiplicative, so that the error
model is additive in logarithms,
\begin{equation}
  \label{eq:errormodel}
  \log \big| y^{cm} \big|
  \;=\;
  \log \big| \mathcal{F}_m^{c}(\thetavec) \big| \;+\; \epsilon^{cm},
  \qquad
  \epsilon^{cm} \sim \mathcal{N}\!\left(0,\ \sigma_{m}^{2}(\thetavec)\right).
\end{equation}
The reference fluxes range over more than an order of magnitude in every
channel, so the likelihood models relative flux errors as Gaussian in log-flux space. The per-channel variance is
decomposed as
\begin{equation}
  \label{eq:koh_split}
  \sigma_{m}^{2}
  \;=\;
  \underbrace{\sigma_{\mathrm{model},m}^{2}}_{\text{inferred}}
  \;+\;
  \underbrace{\sigma_{\mathrm{obs},m}^{2}}_{\text{prescribed}},
\end{equation}
as suggested in \cite{kennedy2001}, with $\sigma_{\mathrm{obs},m}$ being the prescribed noise from the reference data, which is aleatoric in nature, or in other words a noise which no amount of data or model improvement can reduce. The inferred term $\sigma_{\mathrm{model},m}$, i.e., the scale of the discrepancy term, represents
the model-form uncertainty of the ansatz. This is epistemic in origin, since a better
ansatz would reduce it, but entering the likelihood as a stochastic term because at
a fixed ansatz no quantity of further reference data removes it.

The discrepancy term is given a form that is resolved per channel and its scale depends upon the
predicted flux,\footnote{$\mathrm{softplus}(u) = \log(1 + e^{u})$, a smooth
strictly positive map, which keeps $\sigma_{\mathrm{model},m}$ positive for any
$a_m$ and $b_m$.}
\begin{equation}
  \label{eq:hetero}
  \begin{aligned}
  \sigma_{\mathrm{model},m}(\thetavec)
  &\;=\;
  \mathrm{softplus}\!\left( a_m + b_m z^{cm}(\thetavec) \right),\\
  z^{cm} &= \frac{\log|\mathcal{F}^{c}_{m}(\thetavec)| - \ell_m}{s_m},
  \end{aligned}
\end{equation}
in which $z^{cm}$ is the log-prediction standardised by the mean $\ell_m$ and
standard deviation $s_m$ of the observed log fluxes of that channel. Setting
$b_m = 0$ recovers a constant scale, so that (Eq.~\ref{eq:hetero}) generalises the
homoskedastic model rather than replacing it, and $b_m < 0$ states that the
multiplicative error diminishes as the flux rises, and vice versa.
Equation (Eq.~\ref{eq:hetero}) models the \emph{scale} of the residual as a function of
the predicted flux. Being zero-mean in log space by construction, it does not infer
a signed correction to the mean response: what the calibration learns is how large
the departure from the reference typically is, not which way it goes at a given
plasma state. A representation able to correct the ansatz systematically is left to
a later iteration of this work.

The coefficients $\thetavec$ in (Eq.~\ref{eq:errormodel}) carry a parametric uncertainty, epistemic like the discrepancy but, unlike it, contracting as reference cases accumulate.
They receive weakly informative priors
$p(\thetavec)$ centred upon the published values. The priors on the discrepancy term parameters, $p(\mathbf{a})$ and $p(\mathbf{b})$, admit a flux dependence without presuming one. The exact prior forms used are stated in
\ref{app:calib}. From Bayes' theorem \cite{jaynes2003},
with $\mathcal{D} = \{y^{cm}\}$ as the set of observed fluxes,
the posterior form is given as
\begin{equation}
  \label{eq:posterior}
  \begin{aligned}
  \underbrace{p(\thetavec, \mathbf{a}, \mathbf{b} \mid \mathcal{D})}_{\text{posterior}}
  &\;\propto\;
  \underbrace{p(\thetavec)\, p(\mathbf{a})\, p(\mathbf{b})}_{\text{prior}}\\
  &\quad\underbrace{
    \prod_{c}\prod_{m}
    \mathcal{N}\!\left(
      \log|y^{cm}|;\ \log|\mathcal{F}^{c}_{m}(\thetavec)|,\ \sigma_{m}^{2}(\thetavec)
    \right)
  }_{\text{likelihood}} .
  \end{aligned}
\end{equation}
The likelihood assumes conditional independence across reference cases and flux
channels. This is a modeling approximation: correlated residuals, whether between
the three channels of one run or across a parameter scan, would alter posterior
concentration and the joint predictive uncertainty.

All the parameters are sampled jointly. The differentiability noted above extends to
the noise parameters as well as the coefficients, so gradient-based inference is
available in both its variational \cite{blei2017} and its Hamiltonian
\cite{betancourt2017} form. A variational approximation trades the shape of the
posterior for speed, and here that shape is itself what is being reported, so
sampling is preferred: the No-U-Turn sampler \cite{hoffman2014nuts} in \numpyro{}
\cite{phan2019numpyro} is used throughout. \ref{app:calib} gives the sampler
settings and the prior widths.

The formulation is not tied to the integrated flux. Any observable reported by the
reference simulations enters (Eq.~\ref{eq:errormodel}) in similar way.
Conditioning additionally on the wavenumber-resolved spectrum for example would increase the
number of observations per case and would constrain coefficients that an integrated
flux does not separate. Whether it would help in generalization is not explored in this work.

Writing $\boldsymbol{\Theta} = (\thetavec, \mathbf{a}, \mathbf{b})$ for the full
parameter vector, the predictive distribution for a new plasma state $x^{\ast}$ \footnote{A plasma state here is the local parameters of the linear problem at one radius: gradients,
collisionality, safety factor, shear, temperature ratio, flux-surface shape and beta (for electro-magnetic cases). It enters only through $\mathcal{F}_m$.}
is given as
\begin{equation}
  \label{eq:predictive}
  \begin{aligned}
  p\!\bigg{(}\log\big|& y^{\ast}_m\big|  \mid  \mathcal{D}, x^{\ast}\bigg{)}
  \;=\;{}\\[-0.6ex]
  &\quad\int \mathcal{N}\!\left(
      \log\big|y^{\ast}_m\big|;\
      \log\big|\mathcal{F}_m(x^{\ast}, \thetavec)\big|,\
      \sigma_{\mathrm{model},m}^{2}(x^{\ast}, \boldsymbol{\Theta})
  \right) \\
  &\qquad\times\; p\!\left(\boldsymbol{\Theta} \mid \mathcal{D}\right)
  \,\mathrm{d}\boldsymbol{\Theta} ,
  \end{aligned}
\end{equation}
In words, the left-hand side is the distribution of the flux in channel $m$ at a
plasma state outside the reference set, given the database and the calibration
performed upon it. The observational noise of (Eq.~\ref{eq:koh_split}) does not enter here.
The width of (Eq.~\ref{eq:predictive}) is used by downstream calculations, and
it is useful to know which part of it is reducible. Conditioning on
$\boldsymbol{\Theta}$ and applying the law of total variance \cite{gelman2013}, we get
\begin{equation}
  \label{eq:varsplit}
  \begin{aligned}
  \operatorname{Var}\!\left[\log|y^{\ast}_m| \mid \mathcal{D}\right]
  &=
  \underbrace{\operatorname{Var}_{\boldsymbol{\Theta}}\!\left[
      \log|\mathcal{F}_m(x^{\ast}, \thetavec)|\right]}_{\text{parametric}}\\
  &\quad+
  \underbrace{\mathbb{E}_{\boldsymbol{\Theta}}\!\left[
      \sigma_{\mathrm{model},m}^{2}(x^{\ast}, \boldsymbol{\Theta})\right]}_{\text{discrepancy}} .
  \end{aligned}
\end{equation}
The discrepancy is evaluated at the flux predicted for $x^{\ast}$ and is therefore
not a constant inherited from the calibration set. The first term contracts as
reference cases accumulate whereas the second term, the discrepancy term does not diminish
however many are added, so that the two terms target further
nonlinear runs, or a better ansatz. Both terms are evaluated by Monte Carlo over posterior
draws, which at flux level costs little, since the coefficients enter only the
algebraic saturation step and the cached linear solve does not depend upon them.
Exemplarily, the two terms are accordingly reported
separately in Fig.~\ref{fig:parity}.
Note that Sec.~\ref{sec:profiles} propagates the same posterior through the
transport solve itself.

How uncertainty is propagated through a forward model is a developed subject
\cite{sullivan2015,smith2013}, and the choice within it is one of cost against
fidelity. Propagating a single point estimate, say the posterior median, returns
a point prediction; sampling the whole posterior through is faithful but costs an
evaluation per draw; perturbation and spectral methods occupy the ground between.
A surrogate of the solver itself is a further route: neural-network surrogates of
\tglf{} \cite{meneghini2017,cao2025winn} reduce a transport evaluation to a network
pass, and at that cost sampling a posterior through the full posterior becomes feasible.

\subsection{\texorpdfstring{$\satthree$}{SAT3} benchmarks}
\label{sec:results}
\begin{figure*}[htbp!]
  \centering
  \includegraphics[width=\linewidth]%
    {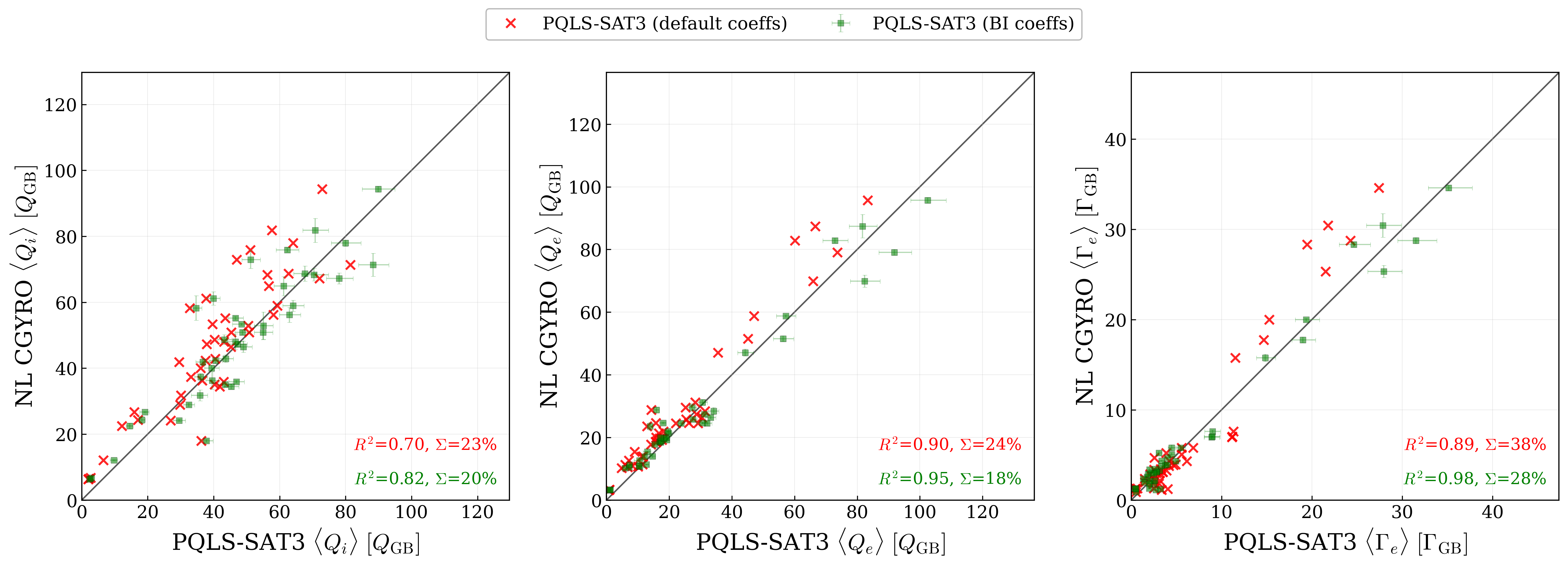}
  \par\smallskip
  {\small (a) coefficient-only intervals.}\label{fig:parity:param}
  \vspace{0.7em}
  \includegraphics[width=\linewidth]%
    {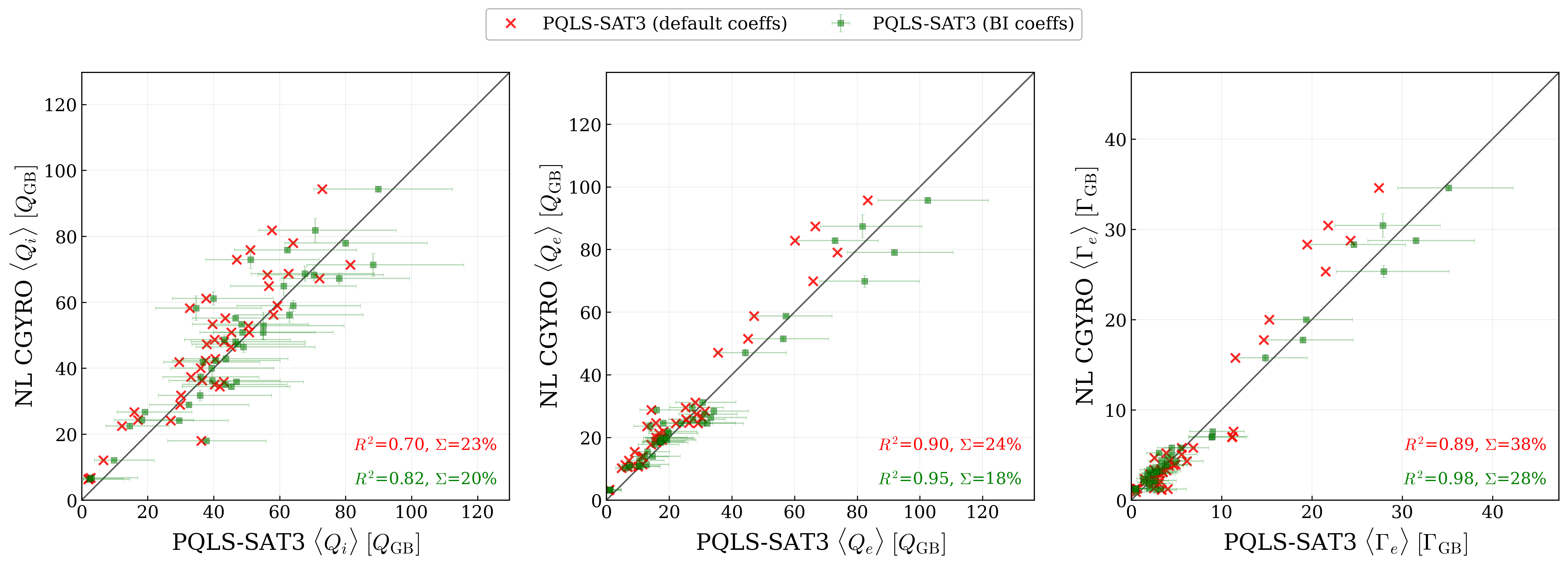}
  \par\smallskip
  {\small (b) coefficient plus discrepancy intervals.}\label{fig:parity:full}
  \caption{Integrated-flux against nonlinear \cgyro{} for the rule closed
    on \pqls{} weights, at published coefficients (red crosses) and calibrated
    (green squares). Columns are ion energy, electron
    energy and electron particle flux. Horizontal bars are $5$ to $95\%$
    predictive intervals and vertical bars the \cgyro{} observed noise. The two rows are
    the two terms of (Eq.~\ref{eq:varsplit}) shown separately, the parametric term
    above and parametric plus discrepancy below, both from the single fit of
    (Eq.~\ref{eq:posterior}).}
  \label{fig:parity}
\end{figure*}

The methodology of Sec.~\ref{sec:calibration} is demonstrated on the $\satthree$
rule \cite{dudding2022}, whose ansatz and calibrated coefficients are summarised in \ref{app:rule}. The reference data are the 43 nonlinear \cgyro{} cases used to calibrate that
rule \cite{dudding2022, dudding2024data}, each indexed by the plasma state defined in
Sec.~\ref{sec:calibration}, Gyrobohm normalised and time-averaged, whose scatter is the prescribed noise
$\sigma_{\mathrm{obs},m}$ of (Eq.~\ref{eq:koh_split}) and the vertical bars of
Fig.~\ref{fig:parity}. The linear drive is \pqls{}, its quasilinear weights being those of
(Eq.~\ref{eq:ql_weight}), computed with the parallel coordinate discretised by
finite differences at $\left(n_{k_x},n_z\right)=(64,32)$ and the moment resolution
$\left(n_u,n_e\right)=(20,10)$. This resolution was settled upon by requiring the
quasilinear fluxes to reproduce the nonlinear reference, so that the residual the
calibration then sees belongs to the saturation rule rather than to an
under-resolved linear solve. The priors and sampler settings used throughout
are given in \ref{app:calib}.

\begin{figure*}[htbp!]
  \centering
  \includegraphics[width=\linewidth]{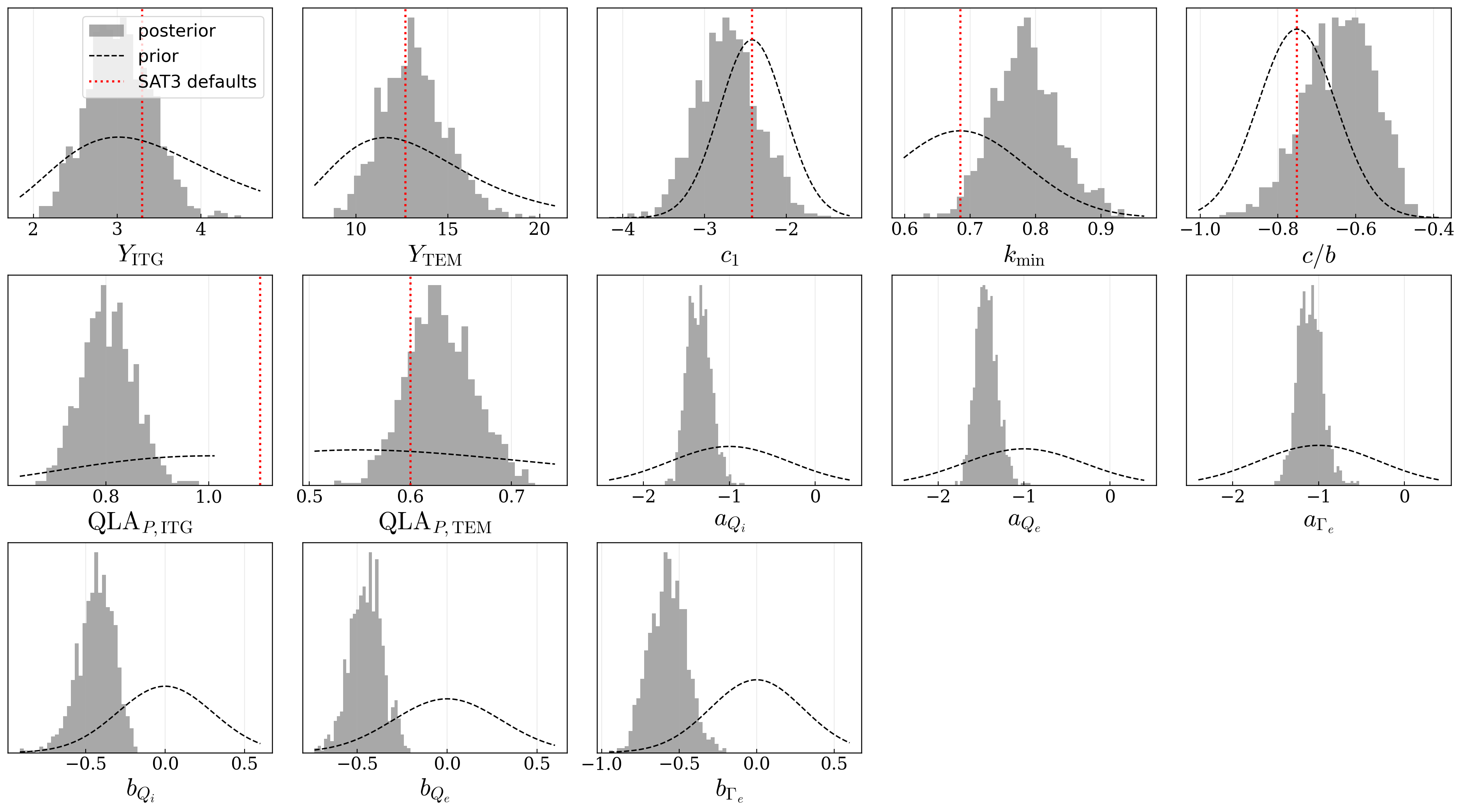}
  \caption{Prior (dashed) against posterior (grey) for the $\satthree$ coefficients and
    the noise parameters of (Eq.~\ref{eq:hetero}), with the published values
    marked (red dotted) where they exist.}
  \label{fig:coeffs}
\end{figure*}

Table~\ref{tab:posterior} and Fig.~\ref{fig:coeffs} give the joint posterior. Some
coefficients are displaced from their priors and sharply identified by the data,
others are left close to where they started, as Fig.~\ref{fig:coeffs} shows case by
case. The slope $b$ is negative in every posterior draw of all three channels, so the flux
dependence of the error is identified rather than merely permitted.

\begin{table}[htbp]
  \centering
  \caption{Posterior mean $\pm$ standard deviation for the calibrated
    $\satthree$ coefficients against
    the published point estimates \cite{dudding2022} and the inferred discrepancy-term parameters of (Eq.~\ref{eq:hetero}). The role of each
    coefficient is tabulated in Tbl.~\ref{tab:coeffroles} of \ref{app:rule}.}
  \label{tab:posterior}
  \small
  \begin{tabular}{@{}lcc@{}}
    \toprule
    Coefficient & Published & Posterior \\
    \midrule
    $Y_{\mathrm{ITG}}$          & 3.3      & $3.03 \pm 0.38$ \\
    $Y_{\mathrm{TEM}}$          & 12.7     & $13.16 \pm 1.76$ \\
    $c_1$                       & $-2.42$  & $-2.71 \pm 0.38$ \\
    $k_{\min}/k_{\max}$         & 0.685    & $0.786 \pm 0.050$ \\
    $c/b$                       & $-0.751$ & $-0.644 \pm 0.090$ \\
    $\qla_{P,\mathrm{ITG}}$ & 1.1 & $0.80 \pm 0.05$ \\
    $\qla_{P,\mathrm{TEM}}$     & 0.6      & $0.63 \pm 0.03$ \\
    \addlinespace
    $a$ \; ($\qi$, $\qe$, $\gpart$)
      & n/a & $-1.35 \pm 0.14$ / $-1.44 \pm 0.13$ / $-1.10 \pm 0.14$ \\
    $b$ \; ($\qi$, $\qe$, $\gpart$)
      & n/a & $-0.44 \pm 0.11$ / $-0.45 \pm 0.09$ / $-0.57 \pm 0.12$ \\
    \bottomrule
  \end{tabular}
\end{table}


Fig.~\ref{fig:parity} compares the fluxes from the calibrated rule against the reference nonlinear data, each panel
reporting $R^2$ and $\Sigma$. Both use the integrated fluxes in Gyrobohm
units with the posterior median as the point prediction: $\Sigma$ is the mean of
$|\mathcal{F}_m - y|/|y|$ over cases, the average absolute percentage error of
\cite{dudding2022}, and $R^2$ is the residual coefficient of determination
$1 - \sum (y-\mathcal{F}_m)^2 / \sum (y-\bar{y})^2$. It can be observed that the recalibration improves on the published
coefficients in every channel by both measures. The two rows of Fig.~\ref{fig:parity} share their medians and differ only in the
width of the horizontal bars. Above, where only the parametric term of
(Eq.~\ref{eq:varsplit}) enters, the intervals are narrow, and wherever the rule
departs from the reference they are too short to bridge the gap. Below, with the discrepancy
added, they widen enough to cover the diagonal in all but a few cases. This is the
concrete form of the statement that a band built from the parametric term alone is
not the predictive uncertainty.

Between the tenth and ninetieth percentiles of the observed flux the discrepancy of
(Eq.~\ref{eq:hetero}) falls from $0.36$, $0.31$ and $0.52$ to $0.16$, $0.12$ and
$0.13$ in $\qi$, $\qe$ and $\gpart$ respectively, the particle channel carrying the
largest slope $b$ and the steepest fall. \ref{app:calib} compares this form against the
homoskedastic case $b_m = 0$. The
flux-dependent discrepancy predicts better without loss of point accuracy (see Tbl.~\ref{tab:sigma_split}).

\section{Profile prediction}
\label{sec:profiles}
\begin{figure}[htpb!]
  \centering
  \includegraphics[width=\linewidth]{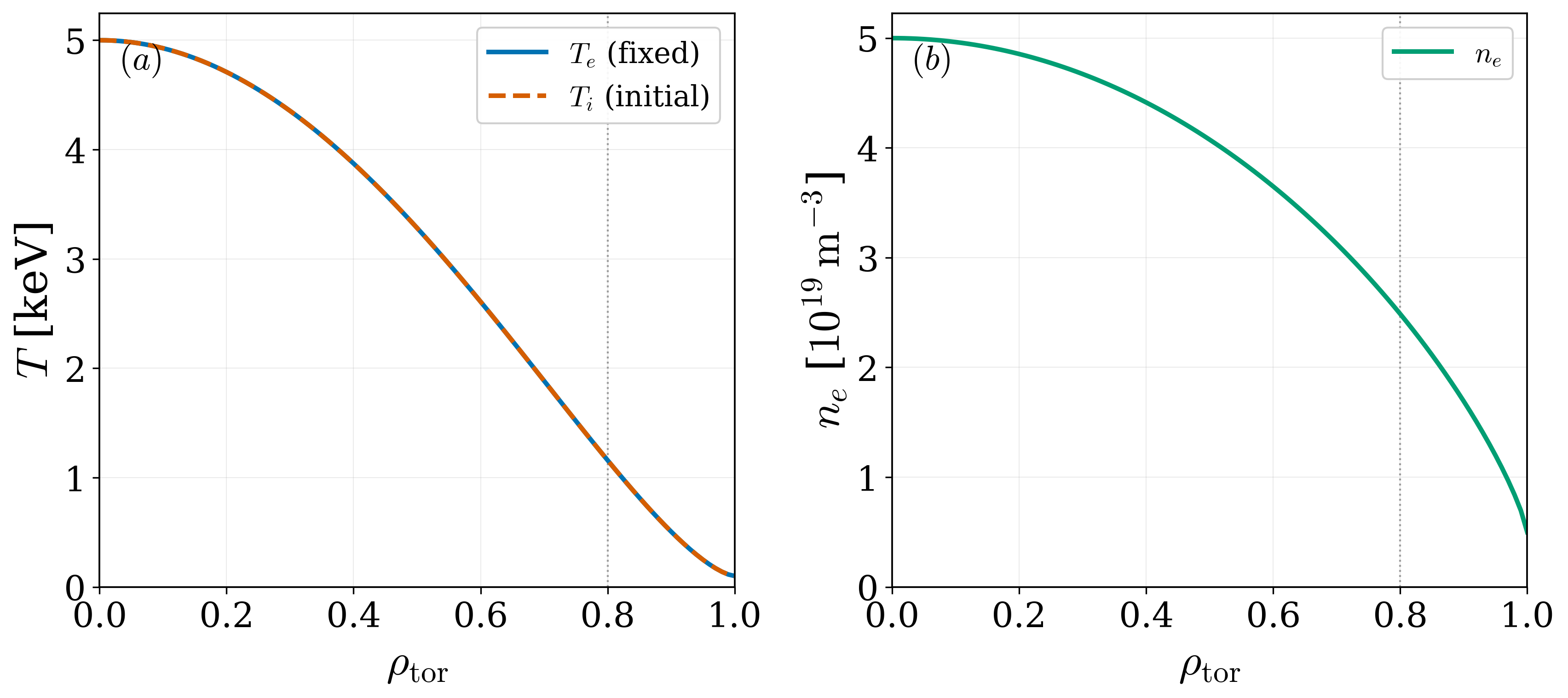}
  \caption{Prescribed electron temperature and initial ion temperature (a) and
    prescribed electron density (b), the dotted line marking the boundary radius
    $\rho_{\rm tor}=0.8$.}
  \label{fig:initialprofiles}
\end{figure}

  We now demonstrate \pqls{} together with the calibration carried out in
  Sec.~\ref{sec:results} on a steady-state transport calculation. The
  calibration constrains the fluxes, whereas a transport calculation is judged by the
  profiles it returns.
  We consider a circular tokamak with $R_0=3$~m, $a=1$~m,
  $B_{\rm axis}=2.5$~T, and $q=1.5+0.8\rho_{\rm tor}^2$. The plasma consists of a
  single hydrogen species, while the density and electron temperature
  profiles are prescribed. Electrons are treated adiabatically in the
  gyrokinetic calculation, and only the ion-temperature profile is varied.
  The prescribed electron temperature nevertheless enters the collisional
  energy exchange between electrons and ions
  \begin{equation}
   \left\langle P_{ei}\right\rangle
   =
   c_{\rm ex}
   \frac{Z_i^2 n_i m_i^{1/2}}
        {\left(m_eT_i+m_iT_e\right)^{3/2}}
   n_e\ln\Lambda\,
   \left(T_e-T_i\right)
   \label{eq:collisional_exchange}
  \end{equation}
  as ion heat source, where $Z_i=1$ and $n_i=n_e$ for the pure-hydrogen plasma. The outer boundary value $T_i(\rho=0.8)$ is held fixed. Figure~\ref{fig:initialprofiles} shows the prescribed profiles together with the ion temperature the solve starts from.

  Using a flux-matching formulation \cite{portals}, we determine the
  steady-state ion-temperature profile by solving a nonlinear optimisation
  problem. The optimisation variables are
  the normalised ion temperature gradients at $N_r=6$ matching radii between
  $\rho_{\rm tor}=0.3$ and $0.8$,
  \begin{equation}
   \mathbf{g}_i
   =
   \left(
   \frac{a}{L_{T_i}}(\rho_1),\ldots,
   \frac{a}{L_{T_i}}(\rho_{N_r})
   \right),
   \label{eq:profile_variables}
  \end{equation}
  from which the full ion temperature profile can be reconstructed.

  At each matching radius, the turbulent heat flux is converted into a
  transported ion power,
  \begin{equation}
   P_{i,j}^{\rm turb}(\mathbf{g}_i,\boldsymbol{\Theta})
   =
   V'(\rho_j)\,
   Q_i^{\rm turb}
   \left(\rho_j;\mathbf{g}_i,\boldsymbol{\Theta}\right),
   \label{eq:turbulent_power}
  \end{equation}
  where $V(\rho_{\rm tor})$ is the volume enclosed by the flux surface,
  $V'=\mathrm{d}V/\mathrm{d}\rho_{\rm tor}$, and $\boldsymbol{\Theta}$ denotes the
  coefficients of the calibrated closure. The corresponding target is the
  cumulative collisional power transferred from electrons to ions,
  \begin{equation}
   P_{i,j}^{\rm src}(\mathbf{g}_i)
   =
   \int_0^{\rho_j}
   V'(\rho)\,
   \left\langle P_{ei}\right\rangle
   \left(\rho;\mathbf{g}_i\right)
   \mathrm{d}\rho .
   \label{eq:source_power}
  \end{equation}
  
  The optimal gradients are obtained by minimising the summed power mismatch,
  \begin{equation}
   \mathbf{g}_i^\ast(\boldsymbol{\Theta})
   =
   \underset{\mathbf{g}_i}{\operatorname{arg\,min}}
   \sum_{j=1}^{N_r}
   \left[
   P_{i,j}^{\rm turb}(\mathbf{g}_i,\boldsymbol{\Theta})
   -
   P_{i,j}^{\rm src}(\mathbf{g}_i)
   \right]^2 .
   \label{eq:profile_optimisation}
  \end{equation}
The coefficients of $\satthree$ are the posterior mean of the
calibration of Sec.~\ref{sec:results}. The band
is the forward propagation of (Eq.~\ref{eq:predictive}) through the flux match, computed by
quasi-Monte Carlo (QMC) using $64$ Sobol draws as \ref{app:delta} describes.

\begin{figure*}[t!]
  \centering
  \includegraphics[width=0.9\linewidth]{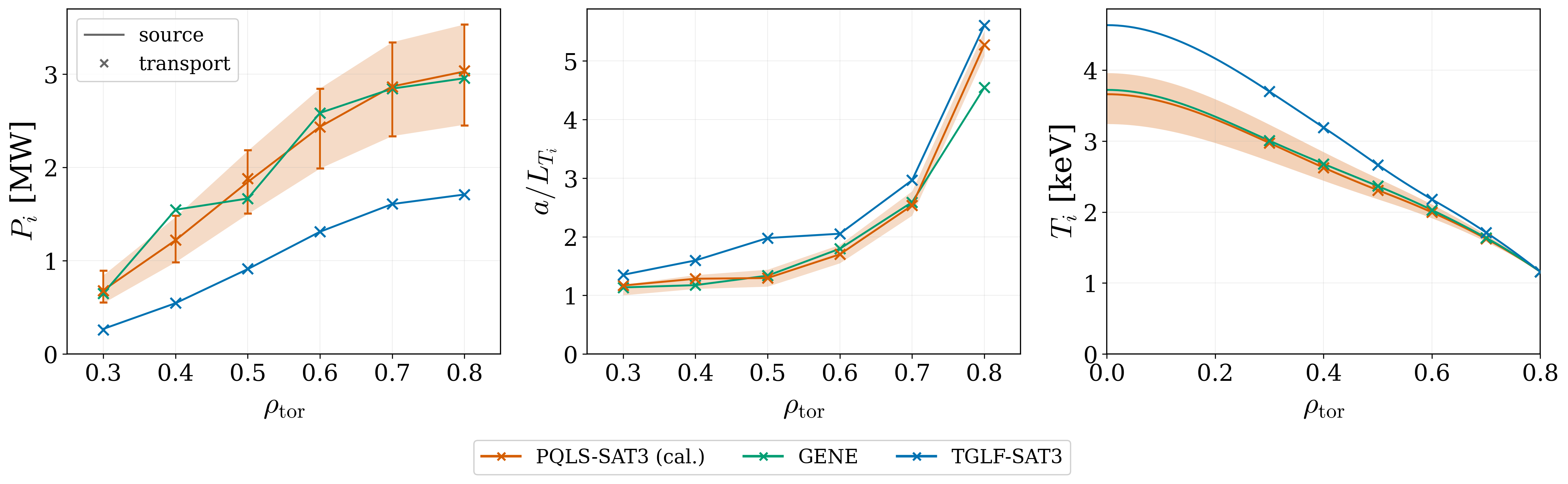}
  \caption{Profile prediction for the adiabatic-electron circular case, $\pm 2\sigma$
    throughout. \pqls{} closed by the calibrated $\satthree$ rule is orange,
    \tglf{} closed by $\satthree$ at its default coefficients is blue, and the
    \gene{} reference is green. Left: the integrated ion power at the six
    flux-matched radii. Centre: the normalised ion temperature gradient
    $a/L_{T_i}$. Right:
    the predicted $T_i$, the band being the quasi-Monte-Carlo envelope of $64$ Sobol
    samples.}
  \label{fig:profileuq}
\end{figure*}

\pqls{} evaluates the quasilinear spectrum on $12$ logarithmically spaced
  binormal wavenumbers covering $0.05 \le k_y\rho_s \le 2.0$ over a single poloidal
  turn, at the same resolution as the calibration of Sec.~\ref{sec:results}.
  The reference is obtained by running nonlinear
  \gene{} simulations with adiabatic electrons in a flux tube of one poloidal turn with
  the perpendicular box $\left(L_x,L_y\right)=\left(128,\,2\pi/k_{y,\mathrm{min}}\right)\rho_s
  \simeq (128,126)\,\rho_s$, $k_{y,\mathrm{min}}\rho_s = 0.05$, and the numerical
  resolutions $\left(n_{k_x},\allowbreak n_{k_y},\allowbreak n_z,\allowbreak n_{v_\parallel},\allowbreak n_\mu\right)
  = (128,\allowbreak48,\allowbreak24,\allowbreak24,\allowbreak12)$
  with the velocity-space extents $L_{v_\parallel} = 3\,v_{\mathrm{th}}$ and
  $L_\mu = 9\,T/B$.

Fig.~\ref{fig:profileuq} gives the result. From the left panel, one can observe that at each matched radius the ion power carried by the turbulence equals the
power delivered by collisional exchange, and the solve adjusts the temperature until
the two agree. Once the saturation rule carries an uncertainty, the flux balance
is not a curve, but a band. The transported power inherits the posterior directly, while
the source inherits it through the profile, since the exchange between electrons and
ions is itself set by the temperature the match returns. In the middle panel, the
normalized temperature gradient predicted by \pqls{} with the calibrated
$\satthree$ rule agrees closely with the \gene{} reference at every matched radius,
whereas the gradient obtained with \tglf{} does not. The right panel shows
the same distinction: the predicted temperature band envelops the \gene{} reference
across the profile, while the \tglf{} prediction lies outside it.

\begin{figure}[t!]
  \centering
  \includegraphics[width=0.65\linewidth]{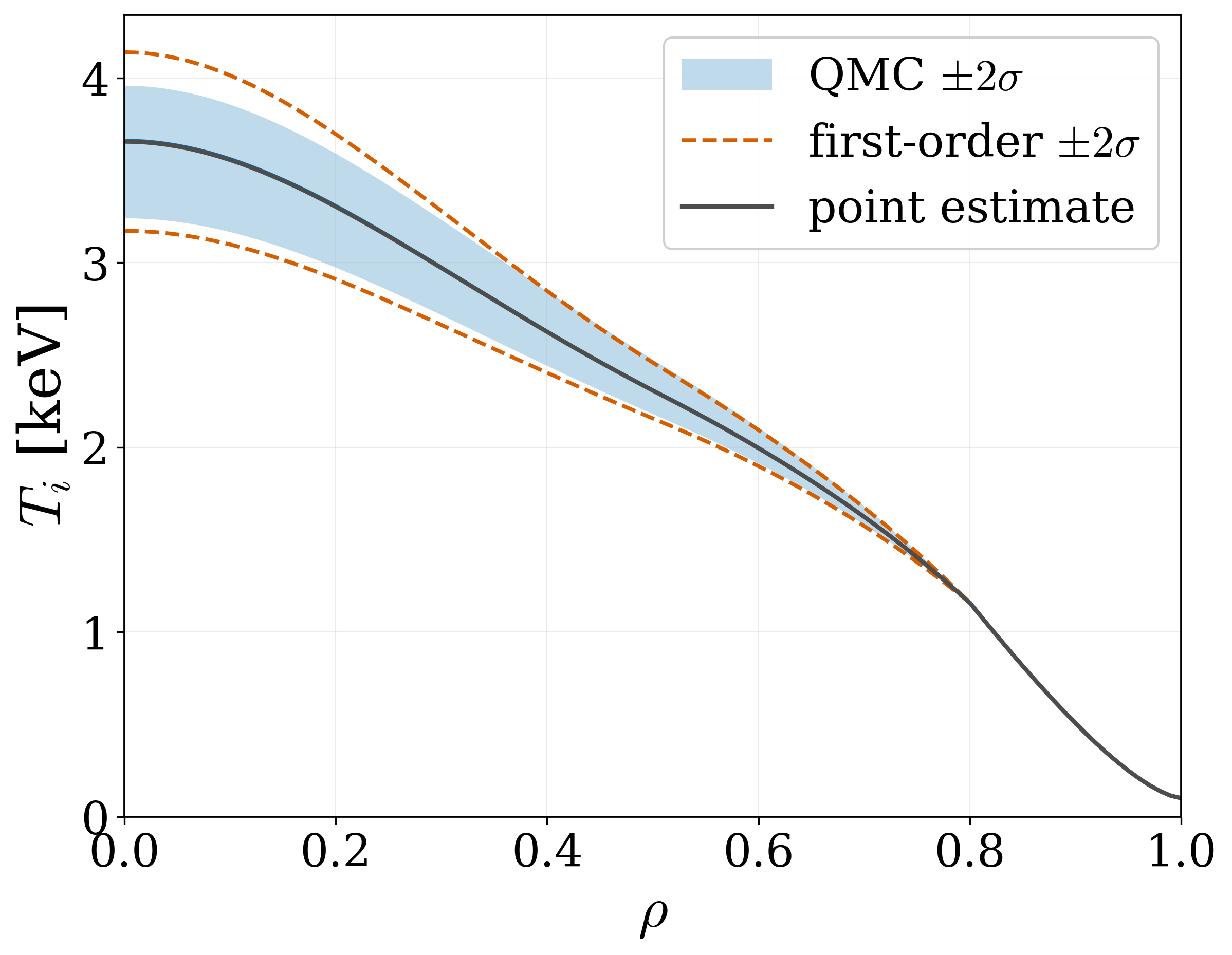}
  \caption{Forward propagation of the coefficient posterior to the ion temperature
    at $\pm 2\sigma$: point estimate (grey), linearised expansion (orange dashed)
    and QMC (blue band).}
  \label{fig:uqcompare}
\end{figure}

Fig.~\ref{fig:profileuq} employs Quasi-Monte-Carlo (QMC) to propagate the uncertainty forward.
QMC converges faster than plain Monte Carlo, $\mathcal{O}(N^{-1})$ against
$\mathcal{O}(N^{-1/2})$. It may still become prohibitive as
the physics complexity increases, since every draw costs a converged flux match at
the resolution that physics demands. Other routes are then needed, as discussed in Sec.~\ref{sec:calibration}. Fig.~\ref{fig:uqcompare} compares three of them: the point
estimate at a single solution, which returns the center line and no band; the
linearized expansion (see \ref{app:delta}), a baseline and one solution per stochastic direction, nine in
all for this case; and QMC at sixty-four. The sampled band is visibly asymmetric,
a skew the linearized route cannot represent by construction. We remark that the magnitude of the asymmetry is a
property of this case rather than a guaranty, since a first-order expansion
carries no information about the curvature it discards.

\section{Conclusions}
\label{sec:conclusions}

We have introduced \pqls{}, a differentiable quasilinear gyrokinetic
transport solver formulated in general magnetic geometry. Its direct
eigenvalue formulation provides access to dominant and subdominant modes at a
user-selected Hermite-Laguerre resolution. Benchmarks against \gene{} show
good agreement in growth rates, real frequencies, and eigenfunctions for the
Cyclone Base Case with adiabatic and kinetic electrons. In the
kinetic-electron case, \pqls{} resolves the coexistence and exchange of
dominance between the ITG and TEM branches. The collisionless microtearing mode
benchmark further verifies the electromagnetic implementation and reproduces
the characteristic parity of the electrostatic and parallel
vector-potential eigenfunctions.

We have also formulated the saturation-rule closure as a Bayesian inference
problem. Rather than determining a single best-fitting parameter set, the
method infers a joint posterior over the saturation-rule coefficients and a
model-discrepancy term to correct for the saturation rule ansatz. Calibrating
the $\satthree{}$ rule on \pqls{} quasilinear weights using nonlinear
\cgyro{} data improves the ion heat, electron heat, and electron particle
flux predictions relative to the published coefficients.

Propagating the posterior through a steady-state transport calculation turns
this flux-level uncertainty into a profile-level prediction. For the circular
adiabatic-electron case considered here, the resulting ion-temperature band
envelops the ion-scale \gene{} reference profile, whereas the prediction
obtained from \tglf{} with the default $\satthree{}$ coefficients lies
outside it. This demonstrates that uncertainty in the saturation closure can
be retained through a complete flux-matching calculation instead of being
discarded at the level of a point estimate. These estimates remain conditional on the calibration database, the priors and the
assumed form of the discrepancy, and agreement on that database does not by itself
establish accuracy at plasma states outside it.

Although \pqls{} supports general magnetic geometry, the saturation rule and
profile application considered here are restricted to tokamaks. Future work
will use nonlinear stellarator databases to develop and calibrate
geometry-appropriate saturation rules, enabling self-consistent stellarator
profile predictions. The access of \pqls{} to multiple eigenmodes will also
allow the influence of subdominant branches on transport to be quantified in
both tokamaks and stellarators in future work. Finally, following the wavenumber-informed
approach developed for \tglf{} in \cite{cao2025winn}, the $k_y$-resolved
outputs of \pqls{} can support physics-informed neural-network surrogates,
with Bayesian active learning reducing the required training data and
accelerating integrated transport calculations. A more expressive discrepancy
model, able to correct mean response as well as its scale, ranging from a Gaussian process on the plasma state to the transformer
architectures now being applied to gyrokinetic turbulence
\cite{paischer2025gyroswin}, will be tackled in the future iteration of the work.

\begin{acknowledgments}
This work was independently funded by Proxima Fusion, and supported by the German Federal Ministry of Research, Technology and Space (BMFTR) under grant number 13F1012A (Joint Project: FUSKI). The simulations were run on the JURECA and JUWELS supercomputers at Juelich, Germany.
\end{acknowledgments}

\section*{Data and code availability}
The data that support the findings of this study are available from the corresponding author upon reasonable request.  \pqls{} and the calibration methodology will be made available upon publication at \url{https://github.com/proximafusion/PQLS.jl}.

\section*{Conflict of Interest}
The authors declare no conflict of interest.

\appendix

\section{Calibration: further details}
\label{app:calib}

\paragraph{Priors} The coefficients are centred upon the published values and the
noise parameters upon no flux dependence,
\begin{equation}
  \label{eq:priors}
  \begin{aligned}
    \log Y_{\mathrm{ITG}} &\sim \mathcal{N}\!\left(\log 3.3,\ 0.3^{2}\right),\\
    \log Y_{\mathrm{TEM}} &\sim \mathcal{N}\!\left(\log 12.7,\ 0.3^{2}\right), \\
    c_{1} &\sim \mathcal{N}\!\left(-2.42,\ 0.4^{2}\right),\\
    k_{\min}/k_{\max} &\sim \mathcal{N}\!\left(0.685,\ 0.1^{2}\right), \\
    c/b &\sim \mathcal{N}\!\left(-0.751,\ 0.1^{2}\right),\\
    \log \qla_{P,\mathrm{ITG}} &\sim \mathcal{N}\!\left(\log 1.1,\ 0.3^{2}\right), \\
    \log \qla_{P,\mathrm{TEM}} &\sim \mathcal{N}\!\left(\log 0.6,\ 0.3^{2}\right),\\
    a_m &\sim \mathcal{N}\!\left(-1.0,\ 0.7^{2}\right), \\
    b_m &\sim \mathcal{N}\!\left(0,\ 0.3^{2}\right).
  \end{aligned}
\end{equation}
The amplitude priors admit a factor of about $1.35$ either side of the published
value at one standard deviation, and about $1.8$ at two. The
level $a_m$ is centred so that the error at typical flux is of order thirty percent, the accuracy such a rule is known to reach against nonlinear data. The slope
$b_m$ carries the same standard deviation, $0.3$, which at one standard deviation
admits a variation of the discrepancy scale of about a factor of three across the
observed flux range; being centred upon zero, it requires a flux dependence to be
demonstrated rather than assumed.

\paragraph{Sampler} Inference uses the No-U-Turn sampler \cite{hoffman2014nuts} in
\numpyro{} \cite{phan2019numpyro}, initialised from a random point at a raised
target acceptance probability, through a \jax{} \cite{jax2018} reimplementation of
the saturation rule that reproduces the reference implementation to machine
precision. The calibration converged with three divergent transitions in a thousand
draws and split $\hat{R}$ within $10^{-3}$ of unity for all parameters
\cite{vehtari2021rhat}; the diagnostic is a single-chain statistic, so it detects a
within-chain trend but not multimodality. A single run supplies both the
coefficients and the noise parameters of Tbl.~\ref{tab:posterior}.

\paragraph{Homoskedastic comparison} Setting $b_m = 0$ in (Eq.~\ref{eq:hetero})
recovers a constant scale. Refitting on the same data gives Tbl.~\ref{tab:sigma_split},
in which the constant scale falls between the two ends of the flux-dependent fit in
every channel, which is what it means to call it an average over the flux range. The
two forms are also compared by the negative log predictive density $\nlpd$, which,
unlike $R^2$ and $\Sigma$, scores a prediction on its centre and its width together
rather than on its median alone, and is therefore the quantity on which an error model
is properly judged. Under (Eq.~\ref{eq:hetero}) the in-sample $\nlpd$ is the lower
in all three channels, so the flux-dependent form attains the better in-sample
predictive density while leaving point accuracy unchanged.

\begin{table*}[htbp]
  \centering
  \caption{The two forms of (Eq.~\ref{eq:hetero}) on the same data.
    $\sigma_{\mathrm{model}}$ is dimensionless, being a scale in log space: constant
    under $b_m = 0$, and evaluated at the tenth and ninetieth percentiles of each
    channel's observed flux under the flux-dependent form. The observational share of
    the residual variance, $\sigma_{\mathrm{obs}}^{2}/\sigma_{m}^{2}$, is given at the
    same two fluxes, with $\sigma_{\mathrm{obs}}$ the median fractional \cgyro{}
    sub-window scatter. $\nlpd$ is in-sample and in log space, lower being better.
    Low and high mean those same two percentiles throughout.}
  \label{tab:sigma_split}
  \setlength{\tabcolsep}{4pt}
  \begin{tabular}{@{}lccccccc@{}}
    \toprule
    & \multicolumn{2}{c}{$b_m = 0$} & \multicolumn{3}{c}{(Eq.~\ref{eq:hetero})}
    & \multicolumn{2}{c}{obs.\ share} \\
    \cmidrule(lr){2-3}\cmidrule(lr){4-6}\cmidrule(lr){7-8}
    Channel & $\sigma_{\mathrm{model}}$ & $\nlpd$
    & $\sigma_{\mathrm{model}}$ low & $\sigma_{\mathrm{model}}$ high & $\nlpd$
    & low & high \\
    \midrule
    $\qi$    & 0.32 & $+0.254$ & 0.36 & 0.16 & $-0.047$
      & $0.6\%$ & $2.9\%$ \\
    $\qe$    & 0.31 & $+0.229$ & 0.31 & 0.12 & $-0.097$
      & $0.8\%$ & $4.4\%$ \\
    $\gpart$ & 0.40 & $+0.485$ & 0.52 & 0.13 & $+0.195$
      & $0.5\%$ & $7.1\%$ \\
    \bottomrule
  \end{tabular}
\end{table*}

\section{The saturation rule and its calibrated coefficients}
\label{app:rule}

The rule calibrated in Sec.~\ref{sec:results} is $\satthree$
\cite{dudding2022}. We do not reproduce its derivation. This appendix fixes the
notation and says what each calibrated coefficient does, see
Table.~\ref{tab:posterior}.

The linear solve supplies, per case, the growth-rate spectrum, the quasilinear
weights $w_m(\ky)$ of (Eq.~\ref{eq:ql_sum}), and the zonal-mixing correction.
From these the rule forms:

\begin{itemize}
  \item $k_{\max}$, the wavenumber at which the zonal-mixing-corrected growth
    rate peaks, and $\gamma_{\max}$ the growth rate there. The zonal-mixing
    velocity is $v_{\mathrm{zf}} = \gamma_{\max}/k_{\max}$.
  \item $x = |w_e / \sum_i w_i|$ evaluated at $k_{\max}$, the mode-character
    ratio, which interpolates the saturation amplitude between the
    ion-temperature-gradient and trapped-electron-mode limits.
  \item $k_{\min} = (k_{\min}/k_{\max})\,k_{\max}$, the lower knee of the
    saturated spectrum.
\end{itemize}

The saturated intensity is built from a quadratic kernel in $\ky$,
\begin{equation}
  \label{eq:sat3_kernel}
  \begin{aligned}
  \sigma_k(\ky) &\;\propto\; \frac{a}{b}\,\ky^{2} + \ky + \frac{c}{b},\\
  \frac{a}{b} &= -\frac{1}{2 k_{\min}},\\
  \frac{c}{b} &= (c/b)\,k_{\max},
  \end{aligned}
\end{equation}
normalised at a reference wavenumber below $k_{\min}$ and raised to a power,
$I(\ky) \propto Y\,\sigma_k(\ky)^{c_1}$, up to factors independent of
$\thetavec$. Table~\ref{tab:coeffroles} lists the seven coefficients calibrated
here: the five saturation-spectrum coefficients of \cite{dudding2022} together with
the two particle quasilinear-approximation endpoints. Here all seven are estimated
jointly under one likelihood.

Each channel's flux carries a further multiplicative factor, the quasilinear
approximation or QLA, interpolated in $x$ between an ion-temperature-gradient
and a trapped-electron-mode value. For the particle channel those two values
are $\qla_{P,\mathrm{ITG}}$ and $\qla_{P,\mathrm{TEM}}$, which are the labels
Fig.~\ref{fig:coeffs} uses.

\begin{table*}[htbp]
  \centering
  \caption{Role of each calibrated coefficient. Published values from
    \cite{dudding2022}; posteriors in Tbl.~\ref{tab:posterior}.}
  \label{tab:coeffroles}
  \setlength{\tabcolsep}{4pt}
  \begin{tabular}{@{}llp{6.8cm}@{}}
    \toprule
    Coefficient & Published & Role \\
    \midrule
    $Y_{\mathrm{ITG}}$  & 3.3      & saturation amplitude in the
      ion-temperature-gradient limit, entering as
      $Y_{\mathrm{ITG}}\,\gamma_{\max}^{2}/k_{\max}^{5}$ \\
    $Y_{\mathrm{TEM}}$  & 12.7     & saturation amplitude in the
      trapped-electron-mode limit, entering as
      $Y_{\mathrm{TEM}}\,\gamma_{\max}^{2}/k_{\max}^{4}$. The different
      $k_{\max}$ power is the origin of the rule's isotope scaling \\
    $c_1$               & $-2.42$  & exponent of the spectral kernel, controls the rate at which $I(\ky)$ decays away from the peak \\
    $k_{\min}/k_{\max}$ & 0.685    & position of the lower knee, hence the
      width of the saturated spectrum \\
    $c/b$               & $-0.751$ & constant term of the quadratic kernel, in
      units of $k_{\max}$ \\
    $\qla_{P,\mathrm{ITG}}$ & 1.1  & quasilinear-approximation factor multiplying
      the \emph{particle} flux, at the ion-temperature-gradient endpoint of the
      mode-character interpolation \\
    $\qla_{P,\mathrm{TEM}}$ & 0.6  & the same factor at the
      trapped-electron-mode endpoint \\
    \bottomrule
  \end{tabular}
\end{table*}

\section{Forward propagation of uncertainties to the profile band}
\label{app:delta}

Sec.~\ref{sec:profiles} propagates the predictive distribution of
(Eq.~\ref{eq:predictive}) through the flux match. Write
$\boldsymbol{\Theta} = (\thetavec, \mathbf{a}, \mathbf{b})$ for the inferred
parameters and let $f(\boldsymbol{\Theta})$ be any scalar the converged solve of
(Eq.~\ref{eq:profile_optimisation}) returns: a profile value at one radius, a
normalised gradient, or an integrated quantity such as the stored energy. Monte
Carlo estimates the distribution of $f$ by sampling the posterior and solving the
flux match once per draw,
\begin{equation}
  \label{eq:mc}
  \boldsymbol{\Theta}^{(i)} \sim p\!\left(\boldsymbol{\Theta} \mid \mathcal{D}\right),
  \qquad
  f^{(i)} = f\!\left(\boldsymbol{\Theta}^{(i)}\right),
  \qquad
  i = 1,\dots,N,
\end{equation}
the reported band being a pair of percentiles of the sample $\{f^{(i)}\}$.

Quasi-Monte Carlo changes only how the draws are placed. A deterministic
low-discrepancy Sobol sequence replaces the independent samples, so that the
parameter space is covered more evenly at the same number of solves.
Sec.~\ref{sec:profiles} uses $64$ such points and takes the band as the $2.3$ to
$97.7$ percentile envelope. Its width is converged in that number: on-axis $T_i$
half-widths of $7.5$, $9.0$, $9.9$ and $9.8$ per cent for the first $8$, $16$, $32$
and $48$ samples, against $9.8$ for the full set.

Every draw can be embarrassingly parallelized, and costs a converged flux match. Where that is unaffordable, $f$ may instead
be expanded to first order about the posterior mean, giving
\begin{equation}
  \label{eq:delta}
  \operatorname{Var}[f]
  \;\simeq\;
  \underbrace{\nabla_{\thetavec} f^{\top}\,
    \boldsymbol{\Sigma}_{\thetavec}\, \nabla_{\thetavec} f}_{\text{parametric}}
  \;+\;
  \underbrace{\sum_{c}
    \left( \frac{\partial f}{\partial \ln m_c} \right)^{2}
    \sigma_{\mathrm{model},c}^{2}}_{\text{discrepancy}} ,
\end{equation}
which is (Eq.~\ref{eq:varsplit}) evaluated at the level of the profile rather than
the flux, with $\boldsymbol{\Sigma}_{\thetavec}$ the posterior covariance of the
coefficients, the discrepancy parameters entering not through this gradient but
through $m_c = \exp(\epsilon_c)$, the multiplicative offset by which the discrepancy
of channel $c$ acts. The derivatives are of the converged solve
rather than of \pqls{} alone: the profile responds only through the minimisation of
(Eq.~\ref{eq:profile_optimisation}), so each is taken by a one-sided difference
between two converged matches, one solve per direction. That costs nine solves here
against sixty-four, and it returns the split between the two terms of
(Eq.~\ref{eq:varsplit}) that a percentile envelope cannot supply. What it does not
return is the shape: the coefficients act on the saturated flux essentially
multiplicatively, and it is that multiplicative structure which makes the sampled
distribution asymmetric, an asymmetry no first-order term can carry.

\bibliography{references}

\end{document}